\documentclass[11pt,draftclsnofoot,onecolumn]{IEEEtran}
\usepackage{cite,amsmath,graphicx,array,bm,amsfonts,amssymb,epsfig,caption}
\def\cS{\mathcal{S}}

\def\cN{\mathcal{N}}
\def\by{\mathbf{y}}
\def\bI{\mathbf{I}}
\def\bQ{\mathbf{Q}}
\def\bU{\mathbf{U}}
\def\bw{\mathbf{w}}
\def\bH{\mathbf{H}}
\def\bY{\mathbf{Y}}
\def\bX{\mathbf{X}}
\def\bA{\mathbf{A}}
\def\bW{\mathbf{W}}

\def\bC{\mathbf{C}}
\def\bq{\mathbf{q}}
\def\bh{\mathbf{h}}
\def\bc{\mathbf{c}}
\def\bx{\mathbf{x}}
\def\bz{\mathbf{z}}

\allowdisplaybreaks

\begin{document}
\title{IoT Random Access in Massive MIMO: Exploiting Diversity in Sensing Matrices}

\author{Mohammad Naseri Tehrani and Shahrokh Farahmand% <-this % stops a space
\thanks{M. N. Tehrani is with the
Institute for Digital Communications, Faculty of Engineering,
Friedrich-Alexander-Universitat (FAU), Erlangen-Nurnberg, Germany, e-mail: \{moh.naseritehrani@fau.de\}.}
\thanks{S. Farahmand is with the Department
of Electrical Engineering, Iran University of Science and Technology, Tehran,
Iran, e-mail: \{shahrokhf@iust.ac.ir\}.}  % <-this % stops a space
\thanks{Parts of this manuscript will be submitted to the International Conference on Communication (ICC) 2020.}}


\markboth{IEEE Transactions on Wireless Communications (submitted)}%
{IEEE Transactions on Wireless Communications (submitted)}



\maketitle

\begin{abstract}
Recently, non-orthogonal codes have been advocated for IoT massive access. Activity detection has been demonstrated to entail common support recovery in a jointly sparse multiple measurement vector (MMV) problem and MMV algorithms have been successfully applied offering various degrees of complexity-performance trade-off. Targeting the small measurement per antenna but large number of antennas setup, independent sensing matrices do offer significant performance advantages. Unfortunately, the IoT random access problem can not readily benefit from this concept as code matrix is fixed over all receiving antennas. Our contributions towards addressing this challenge are as follows. First, independent small-scale fading across antennas and users is established as a possible source of sensing matrix de-correlation. Secondly, two novel algorithms are proposed which exploit this partial de-correlation and collect sensing matrix diversity. Enjoying a low-complexity, these methods do offer great practical advantages as they target small measurement size, which is indeed severely constrained due to limited coherence time/bandwidth, but instead compensate for it by using a large array of antennas. Thirdly, probability of failure (PoF) for these methods are rigorously derived and corresponding measurement inequalities are presented. Fourthly, extensive simulations are conducted to confirm the superior performance of these methods versus state of the art.
\end{abstract}

%% Note that keywords are not normally used for peerreview papers.
%\begin{IEEEkeywords}
%Sparsity, Support Recovery, Massive MIMO, On-Off Channel, Random Access, Internet %of Things (IoT).
%\end{IEEEkeywords}

% \fi
\IEEEpeerreviewmaketitle

\section{Introduction}
It is well-known that multi-user massive MIMO (mMIMO) decouples the channel between the base station (BS) and different user terminals (UTs) into single-user deterministic channels free from fast-fading, interference, and noise \cite{mar10}. This comes as a consequence of law of large numbers and is known as channel hardening \cite{commag14}. In order to harvest the promising advantages mMIMO has to offer for internet of things (IoT), the challenging problem of initial random access to a mMIMO BS should be addressed. The challenge is that the number of orthogonal sequences needed for channel estimation are limited and they can not be allocated to UTs on a permanent basis. Therefore, some dynamic allocation is needed on a per demand basis and the process for this random access to pilots (RAP) has been recently investigated \cite{carvalho17}. Some versions of the algorithm are
able to recover data even in the presence of collisions \cite{sorensen14},
\cite{carvalhotwc17}. This scheme is reminiscent of slotted Aloha and has
been considered for 4G LTE random access as well \cite{emil16}.

The orthogonal pilots shortage is even more pronounced for IoT connections as
their numbers is considerably larger. These devices are inactive most of the
time but become active and communicate at a low data rate once in a while. In the extreme case, they transmit only one bit announcing their activity. Accordingly, we consider the so-called on-off random access channel which is a good approximation of IoT devices behavior. This channel has been investigated in the SISO case by  \cite{fletcherarx09}, \cite{bockelmann13}, \cite{access15}.

In the on-off random access channel, users are assigned permanent independent
pseudo-random non-orthogonal codes that exist in abundance. When a particular user
becomes active it transmits its code. Allowing for error-free detection in spite of
interference, is the fact that active users form a small subset of total
users. Hence, exploiting sparsity-enabling techniques, one can recover all
the active users correctly provided that code length is sufficiently large.
This scenario is reminiscent of overloaded CDMA systems but with a low
activity factor \cite{zhu11}. An extension of on-off random access channel
which involved simultaneous activity detection and data decoding was looked
at by \cite{beyene17}, \cite{ghadyani17}, and \cite{Chen18} in the massive MIMO context. These references realized that user activity detection in such channels
amounts to a sparse common support recovery problem for multiple measurement
vectors (MMVs). Subsequently, \cite{beyene17} proposed CoSaMP and MMV
thresholding while \cite{ghadyani17} applied group least absolute shrinkage
and selection operator (group-LASSO) to detect active terminals. Two versions of approximate message passing (AMP) tailored to the specific massive MIMO channel distribution have been advocated by \cite{Chen18}. Certainly, other well-known MMV support recovery algorithms such as simultaneous orthogonal matching pursuit (S-OMP) \cite{Tropp06}, group orthogonal matching pursuit (G-OMP) \cite{lozano09}, multiple Bayesian sparse learning (M-SBL) \cite{Wipf07}, reduce and boost \cite{Mishali08}, and so on can be applied to this problem. However,  all these methods use a fixed sensing matrix for support recovery and therefore their performance is limited when available measurement size is small.

Relevant to our problem, two other general lines of research should be pointed out here. In the first category, it is assumed that local scattering occurs only around the IoT devices and BS is located at a high tower with no scattering nearby. This amounts to reflections arriving to the BS come from a narrow angular spread. Thus, a massive antenna BS can detect angle of arrival as well as localize individual reflections \cite{xie16}. Subsequently, small-scale fading will vanish. Recently, an iterative scheme that alternates between recovery of sparse user activity pattern and recovery of sparse angular spread parameters for individual users has been proposed \cite{schober19}. We should highlight the fact that we have a different model as we allow for scattering near BS as well. Hence, there is no sparsity in angular spread, and small-scale fading exists because individual reflections can not be localized. In addition, \cite{schober19} assumes that active users remain the same during a coherence time, which is different from our system model. Second line of work pertains to sparsity pattern tracking as set forth by \cite{du17}. In this regards, IoT devices do have a low amount of data to transmit. Thus, when they become active they remain active for several random access slots. This correlation in sparsity pattern is tracked by \cite{du17}. We should point out that we assume independent activity across random access slots while our algorithms do tackle unknown CSI, which is assumed known by \cite{du17}. Finally, \cite{du18} assumes an unknown sparsity level and proposes an algorithm that adapts to changing sparsity level over time. Again, \cite{du18} assumes full CSI is available. Contrary to these two works, we assume CSI is not available at first but is acquired over time. Furthermore, none of these works have considered diversity in sensing matrices. They use different criteria also. Both \cite{du17},\cite{du18} consider symbol error rate (SER). In our scenario, devices transmit only one bit when active. Hence, probability of correct support recovery or its complement, which is probability of failure (PoF), is equivalent to the more complicated SER.    

As an example of diversity in sensing matrices, \cite{baron09} has investigated a setup where independent sensing matrices are exploited and non-zero elements of the sparse unknown matrix are independent as well. Then, simple thresholding which they refer to as trivial pursuit (TP) can almost surely recover the true support even with a single measurement per sensor provided that number of sensors, which equals number of antennas in our problem, go to infinity. Exploitation of this remarkable phenomenon for IoT-based random access is missing in the current literature. One should note that the targeted small measurement size but large antenna regime offers great practical advantages as small measurement size, which is limited by coherence time/bandwidth and out of our control, is compensated for by a large number of antennas which we have control over. Our main contributions towards addressing this issue can be enumerated as follows.

\begin{enumerate}
  \item[{\bf 1.}] When a device transmits its code, the same code is received by all antennas. Hence, IoT-based random access is not readily amenable to independent sensing matrices formulation. We offer to exploit independence of small-scale fading across users and antennas to partially de-correlate sensing matrices.
  \item[{\bf 2.}] Based on item {\bf 1}, we propose two novel algorithms, namely opportunistic maximum correlation (OMC) and opportunistic thresholding detector (OTD) that can collect the sensing matrix diversity offered by partial de-correlation with low-complexity. OTD operates only when power control is applied, while OMC can operate satisfactorily even when powers are not equal.
  \item[{\bf 3.}] Leveraging recent concentration of measure results such as Martingale-based bounds, we rigorously derive upper bounds on probability of failure (PoF) for the proposed methods. Furthermore, measurement inequality which determines the rate at which system
  parameters should increase to ensure probability of failure in exact
  support recovery goes to zero are evaluated. It is revealed that OTD maintains its superior performance in this large measurement size setup while OMC loses its edge.
  \item[{\bf 4.}] Complementary to the large measurement regime which is analytically evaluated in item {\bf 3}, we corroborate the improved performance of the proposed algorithms in {\bf 2} versus existing alternatives via comprehensive simulations. Specifically, we demonstrate that when measurements are a few, but number of antennas is large, our simple thresholding algorithms significantly outperform the state of the art.

\end{enumerate}

\subsection{Prior Art on Performance Analysis}
As one of our contributions amounts to rigorous performance analysis, we offer a background on literature that have looked at this issue. Fundamental results for the optimum MMV detector with a fixed sensing matrix can be found in \cite{chen06}, \cite{davies12}, \cite{jin13}, and \cite{tang10}. Maximum correlation and thresholding detectors are considered to be one of
the weakest recovery methods. While \cite{gribonval08} has performed noisy
worst-case and average-case analysis, noiseless average-case analysis has
been carried out by \cite{eldar10}. For the sake of completeness, the
performance of SMV thresholding has been investigated by \cite{fletcher09}.
Provided measurement size is large enough \cite{eldar10} suggests the
probability of error can be driven to zero by increasing the number of
antennas. Given our interest in low measurement size and large number of
antennas, this result provides mixed conclusions. It verifies that increasing
the number of antennas indefinitely can be helpful to thresholding but it
also demands that measurement size is not very small. Unfortunately, this is
only a sufficient condition and does not state if recovery is still possible
when measurement size falls below the given threshold. As a final note, it
should be mentioned that thresholding performance can equal that of the
combinatorial optimal decoder! For example, \cite{reeves09} has proven that
in the average-case analysis and for low SNR, thresholding is the optimum
detector for both complete and partial support recovery criteria.

The aforementioned results pertained to a fixed sensing matrix. If MMSE is selected as criterion, the performance of independent versus
fixed sensing matrices will be almost the same \cite{zhu17}. However,
\cite{park17} proved that measurement inequalities can be improved for
independent sensing matrices compared to a fixed one if probability of
correct support recovery is selected as performance criterion. Unfortunately,
there still exists a nontrivial gap between necessary and sufficient
conditions proposed by \cite{park17}. Specifically, one can let number of
antennas go to infinity and obtain a measurement size as small as one which
is desirable in the necessary condition. However, the sufficient condition
demands that measurement size be greater than the number of active users even
for large number of antennas. Given the result of trivial pursuit, we
conjecture that the necessary condition in \cite{park17} is also sufficient
because the optimal decoder is guaranteed to outperform trivial pursuit and
hence should be able to surpass its performance.

The rest of the paper is organized as follows. Section II presents the
problem formulation. Section III offers intuition on how to de-correlate sensing matrices. Section IV derives the novel OMC and OTD algorithms in detail. Section V offers the main results in performance analysis with proofs relegated to the appendix. Finally, Section VI
provides numerical results and Section VII concludes the paper.

{\bf Notation:} Uppercase boldface letters are used for matrices and
lowercase boldface letters are used for vectors. Calligraphic letters are
used to represent index sets. If $\cS$ is an index set, $\bA_{\cS}$ denotes
the submatrix generated by keeping only those columns of $\bA$ whose index
belongs to $\cS$. Similarly, $\bA^{\cS}$ denotes a submatrix generated by
keeping only rows of $\bA$ corresponding to the index set $\cS$. $\|\bA\|_F$
denotes the Frobenius norm of $\bA$, while $\|\bA\|_2$ denotes the spectral radius, and $\|\bx\|_{\ell}$ represents $\ell$
norm of vector $\bx$. 

\section{Problem Formulation}
Let us consider the uplink of a single-cell with a single base station (BS)
where an $M$-antenna massive MIMO BS serves $N$ single antenna users. In an
IoT scenario, $N$ is large but each user transmits only sporadically. In the
extreme case, which is considered here, each active user transmits a single
bit and inactive users do not transmit at all. It is assumed that $K$ users
out of $N$ are active at each random access slot where $K \ll N$. Note that
we need not know $K$ exactly but an upper bound is sufficient. 

Continuing with our setup, one considers a coherence interval \cite{mar16}
which consists of $T_c$ time slots, where each time slot equals one OFDM
symbol length. In fact, $T_c$ represents the coherence time and $B_c$
denotes the number of OFDM sub-carriers falling into a single coherence
bandwidth. This coherence interval will offer $T_cB_c$ channel uses or a
resource block of length $T_cB_c$ among which channel gain matrix is both
fixed and flat. Typical values for coherence interval in various
scenarios can be found in \cite{mar16}. Now, define $L$ to be the length of
pseudo-random non-orthogonal code which is assigned independently and
permanently to each individual user. We use $\bc_n$ which is a vector of size
$L\times 1$ to represent the code corresponding to user $n$. Here, we chose
Rademacher sequences as codes, where each entry $\bc_n(i)$ takes values
$\{\pm1/\sqrt{L}\}$ equiprobably. The choice of Rademacher codes is made
based on a combination of satisfactory simulation performance and convenience
of mathematical analysis. Setting $L$ equal to a positive integer multiple of $B_c$ proves to be convenient. Upon defining the positive integer $a\geq 1$, one has $L=aB_c$ and $T_s:=\lfloor T_cB_c/L\rfloor=\lfloor T_c/a\rfloor$. Note that the same
random access procedure is utilized $T_s$ times per coherence interval, but
each time a new independent set of $K$ users are active and transmitting.
To determine the active/inactive status of each user an $N\times 1$ vector
$\bq$ is defined whose $n$'th component correspond to user $n$. If user $n$
is active in a particular random access slot, $q_n=\bq(n)=1$ otherwise
$q_n=\bq(n)=0$. As a result, $\bq$ consists of $K$ ones and $N-K$ zeros. In
addition, we consider a block fading model on fast-fading channel gains
meaning that channel is constant for one coherence time and changes
independently afterwards.

Next, define $\bH$ of size $M \times N$ to be the channel gain matrix between
BS and users where its $\{m,n\}$'th entry $H_{m,n}$ represents channel gain
from user $n$ to antenna $m$. It is assumed that $H_{m,n}$s are independent
identically distributed (IID) Gaussian with zero-mean and unit variance. The
path loss is absorbed into the power term to be defined later on. At a
particular random access slot, each active user transmits
$\sqrt{P_{u,n}}\bc_n$ and inactive users remain silent. BS's $m$'th antenna
receives $\forall m=1,\ldots,M$
\begin{equation} \label{rx_single_ant}
\by_m:=\displaystyle \sum_{n=1}^{N} q_n \sqrt{P_n} H_{m,n} \bc_n+\bw_m = \displaystyle \sum_{k\in \cS} \sqrt{P_k} H_{m,k} \bc_k+\bw_m.
\end{equation}
Here, $P_k$ represents the product of the transmit power, path-loss, and
receive antenna gain. It can be thought of as the overall power of user $k$
received at the BS. Set $\cS$ represents the active user set which is of size
$K$. Finally, $\bw_m$ is receiver noise which is ${\cal N}({\bf
0},\sigma_w^2\bI)$. Note that our parameters assume real values as users
transmit only a single bit in the In-phase component and the quadrature
component is not used at all. To rewrite \eqref{rx_single_ant} compactly, we define the following matrices:
\begin{eqnarray}\label{matrices1}
\bY:=\left[\begin{array}{cccc} \by_1 &\vline \quad\by_2 &\vline \quad\cdots &\vline\quad \by_M \end{array}\right]
\in {\bf R}^{L\times M},\qquad \bW:=\left[\begin{array}{cccc} \bw_1 &\vline \quad\bw_2 &\vline \quad\cdots &\vline\quad \bw_M \end{array}\right]
\in {\bf R}^{L\times M}
\end{eqnarray}
where $\bY$ denotes the total received measurements at the BS with
measurements at antenna $m$ appearing at the $m$'th column. Matrix $\bW$
represents the overall noise. Let us also define $\forall m=1,\ldots,M$ the
following variables $\bx_m\in {\bf R}^{N\times 1}$
\begin{eqnarray}\label{matrices2}
\bx_m:=\Big[\begin{array}{cccc}q_1\sqrt{P_1}H_{m,1} & \vline \quad q_2\sqrt{P_2}H_{m,2}&\vline\quad\cdots &
\vline\quad q_N\sqrt{P_N}H_{m,N}\end{array}\Big]^T,
\end{eqnarray}
and
\begin{eqnarray}\label{matrices3}
\bX:=\left[\begin{array}{cccc}\bx_1 &\vline \quad\bx_2 &\vline \quad\cdots &\vline\quad \bx_M \end{array}\right]
\in {\bf R}^{N \times M},\qquad \bC:=\left[\begin{array}{cccc}\bc_1 &\vline \quad\bc_2 &\vline \quad\cdots &\vline\quad \bc_N \end{array}\right]
\in {\bf R}^{L \times N}
\end{eqnarray}
Combining \eqref{matrices1}, \eqref{matrices2}, \eqref{matrices3}, the per
antenna expression in \eqref{rx_single_ant} can be written in compact form as
\begin{equation}\label{main}
\bY=\bC\bX+\bW.
\end{equation}
In sparse recovery nomenclature, $\bC$ is referred to as the sensing matrix.
The first problem is to recover $\bX$ from $\bY$ and $\bC$. Since $\bX$
enjoys a group sparse structure, this problem becomes a well-known case of
multiple measurement vector (MMV) sparse recovery and notable MMV algorithms
have been applied to this problem as was elaborated in the introduction. All these algorithms have well-proven merits in the large parameter regime. However, we are looking for an algorithm that performs best in the other end of the algorithmic spectrum. That is, the algorithm that performs best for the smallest possible $L$ while $M$ is taken arbitrarily large. 

In the single measurement vector (SMV) setup, i.e. $M=1$,
it is well-known that if $L$ increases faster than a certain rate, recovery
error can be made arbitrarily small \cite{wainwright09}. Two main criteria
for recovery error are Mean Square Error (MSE) also known as $\ell_2$-norm
and probability of exact support recovery. Note that the non-zero support of
$\bX$ represents the active users. Hence, in our setup one only needs to
recover the support accurately. As a result, MSE is not a good criterion
because MSE can be small while support is reconstructed incorrectly
\cite{wainwright09}. Therefore, probability of exact support recovery will be
our choice of criterion. Targeting the small $L$ and large $M$ regime, we show in the next section that models with independent sensing matrices perform very well. Then, we allude to our main idea for de-correlating sensing matrices in our target application.

\section{Intuitive Observations}
Two observations are presented here which form the foundation for the
two novel algorithms that will be proposed in the next
section. They both point to the same direction. Specifically, they suggest
that if the constant sensing matrix $\bC$ in \eqref{main} is replaced by IID
sensing matrices $\bC_m$ that is
\begin{equation}\label{main_b}
\by_m=\bC_m\bx_m+\bw_m,\qquad m=1,\ldots,M
\end{equation}
then support recovery performance can be greatly improved.

\subsection{Observation I}
Suppose system model is given by \eqref{main_b} where $\bC_m$'s, which are
sensing matrices, are IID with independent Gaussian entries. Furthermore,
suppose that $\bX:=[~\bx_1 ~|~\bx_2 ~|~\cdots ~|~\bx_M ~]$ enjoys a common
sparse support, i.e., many rows are identically zero, but its nonzero entries
assume values that are IID Gaussian and independent from $\bC_m$'s. Then the
trivial pursuit proposed by \cite{baron09} can recover the support of $\bX$
even with $L=1$ provided that $M$ grows large. Note that trivial pursuit is
nothing but the low-complexity thresholding algorithm applied to the
independent $\bC_m$ case. This result is remarkable in the sense that
with $L=1$ it is impossible to estimate the nonzero entries in $\bX$ even if
one knew the true support. Yet, $L=1$ is sufficient to recover the common
support for large $M$. Finally, it should be noted that the proposed decoder
is not optimal and indeed very simple. 

\subsection{Observation II}
Let us consider the model \eqref{main} but assume that channel gain matrix
$\bH$ is known. Furthermore, choose $L=1$ which means that there is no
pseudo-random codes $\bc_n$ but just a scalar $c_n=\pm 1$. Without loss of
generality, assume all $c_n=+1$. Subsequently, the $\bC$ matrix becomes a row
vector $\bc^T:=[1 ~1 ~1 \cdots 1]^T\in {\bf R}^{1\times n}$ and measurements
per antenna become a scalar. If we stack the scalar measurements per antenna
in a row vector $\by^T:=[y_1,y_2,\ldots,y_M]^T$, we obtain:
\[
\by^T=\bc^T\bX+\bw^T, \qquad ~y_m=\bc^T\bx_m+w_m,~\forall m=1,\ldots,M
\]
Now, if one moves the known channel gains from the unknown $\{\bx_m\}$'s to
the $\bc^T$ vector, the following equivalent equation ensues.
\begin{equation}\label{known_h}
\by=\bH\tilde{\bx}+\bw
\end{equation}
where we have defined $\tilde{\bx}:=\left[\begin{array}{cccc}q_1\sqrt{P_1} & \vline \quad q_2\sqrt{P_2}&\vline\quad\cdots&\vline\quad q_N\sqrt{P_N}\end{array}\right]^T
\in {\bf R}^{N\times 1}$. Note that with $L=1$ and known $\bH$, $\by$ becomes of size $M$ and the
vector $\bx$ which is of size $N$ becomes $K$-sparse. Given that entries of
$\bH$ are independent standard Gaussians, one can apply the known results in
SMV case to conclude that if $M$ grows larger than a certain threshold then
perfect recovery is possible using either the optimal decoder
\cite{wainwright09} or even a simple thresholding algorithm
\cite{fletcherarx09}. Indeed, $\bH$ can be seen as a natural code assigned to
the different users by the environment. Therefore, if we knew $\bH$, we could
have used it to distinguish different users and there was no need for codes
of length $L$. To signify the relationship between observation II and IID
sensing matrices $\bC_m$'s, note that according to \eqref{known_h}, the
$m$'th antenna receives $y_m=\bh^m\tilde{\bx}+w_m$ where $\bh^m$ represents
the $m$'th row of $\bH$. For a given $m$, $\bh^m$ can be thought of as
$\bC_m$ and then we can see that $\bC_m$'s are IID and that is the reason we
obtain such a good performance with small $L$ and large $M$.

\section{Exploiting diversity in sensing matrices}
As witnessed by \eqref{main}, code matrix $\bC$ is fixed across antennas.
However, channel gains are independent across antennas and users. Therefore,
we can de-correlate the code matrix $\bC$ by multiplying it with the channel
gain matrix $\bH$. Unfortunately, we do not have access to the true $\bH$
values. So, we replace the true channel gains with their estimates. This way, we
partially de-correlate the sensing matrices. As the algorithm proceeds in time, more users become active and hence their channel estimates will be
available. Therefore, more de-correlation occurs. Note that user terminals' speeds should be lower than a
certain threshold to ensure sufficiently long coherence interval which allows
for complete de-correlation. However, if the coherence interval is not long
enough partial de-correlation is still achieved. Utilizing this idea, we
propose opportunistic maximum correlation (OMC) and opportunistic
thresholding detector (OTD) algorithms. We begin with the simpler OMC first.

\subsection{Opportunistic Maximum Correlation}
Suppose we are at the first random access slot, and we do not know any of the
channel gains. Then, once the measurement matrix $\bY$ is observed, we
perform thresholding on model \eqref{main} as follows. For $n=1,2,\ldots,N$
define the decision statistic $\theta_n$:
\begin{equation}\label{dec_stat}
\theta_n=\frac{1}{M}~\|\bc_n^T\bY\|_2^2=\frac{1}{M}\sum_{m=1}^{M}\left(\bc_n^T\by_m\right)^2
\end{equation}
Then, select the $K$ largest $\theta_n$ values and set their indices to be
the support estimate. Let us refer to this set as $\hat{\cS}$. Then, apply LS
to the over determined problem obtained from keeping only the indices
corresponding to $\hat{\cS}$:
\begin{equation}\label{least-squares}
\bY=\bC_{\hat{\cS}}\bX^{\hat{\cS}}+\bW, \qquad 
\hat{\bX}^{\hat{\cS}}:=\left(\bC_{\hat{\cS}}^T\bC_{\hat{\cS}}\right)^{-1}\bC_{\hat{\cS}}^T\bY
\end{equation}
Once this estimate is obtained, one notes that from \eqref{matrices2},
\eqref{matrices3}, $\bX_{i,j}=q_i\sqrt{P_i}H_{j,i}$. Given that $q_i=1$ for
active users, and assuming known $P_i$'s at the BS, one can estimate the
channels between user $n$ and BS as
\begin{equation} \label{chan_est}
\hat{\bh}_n:=\frac{\left({\hat{\bX}}{^{^{n}}}\right)^T}{\sqrt{P_n}}
\end{equation}
Given such an initial random access slot at the beginning of each coherence
interval, we move on to define OMC for any subsequent random access slots.
Suppose, we are at an arbitrary random access slot greater than one. Let
$\Lambda$ represent the index set of users that have been active at least
once in this coherence interval before the current random access slot. Then,
define the following parameters:
\begin{equation}\label{otd1}
\tilde{H}_{i,j}:=\left\{\begin{array}{ll}\hat{H}_{i,j}, & j\in\Lambda\\
1, & \mbox{O.W.}\\\end{array}\right., \qquad
\check{H}_{i,j}:=\left\{\begin{array}{ll}1, & j\in\Lambda\\
H_{i,j}, & \mbox{O.W.}\\\end{array}\right..
\end{equation}
Upon defining $\forall m=1,\ldots,M$
\begin{equation}\label{otd3}
\check{\bx}_m:=\left[~q_1 \check{H}_{m,1}\sqrt{P_1}~|~ q_2 \check{H}_{m,2}\sqrt{P_2}~|~ \ldots ~| ~q_N \check{H}_{m,N}\sqrt{P_N} ~\right]
\end{equation}
one arrives at the following set of equations $\forall m=1,\ldots,M$ which
are equivalent to \eqref{main}:
\begin{eqnarray}\label{otd4}
\by_m=\bA_m\check{\bx}_m+\bw_m, \qquad \bA_m=\left[~\tilde{H}_{m,1}\bc_1~|~\tilde{H}_{m,2}\bc_2~|~\cdots ~|~\tilde{H}_{m,N}\bc_N~\right].
\end{eqnarray}
Note that compared to the fixed sensing matrix in \eqref{main}, we now have a
set of sensing matrices $\bA_m$ which are partially uncorrelated. Next, we
apply maximum correlation to the model \eqref{otd4}. Let us define the
decision statistics:
\begin{equation}\label{otd5}
\theta_n=\frac{1}{M} \sum_{m=1}^{M} \left(\tilde{H}_{m,n}\bc_n^T\by_m\right)^2
\end{equation}
Then, we select the $K$ largest $\theta_n$'s and choose their indices to
represent the true support. To ensure that active devices are not missed from
the support, we add the active devices found by the ordinary thresholding
applied to \eqref{main} and refer to the union of these set as $\hat{\cS}$
and use the LS estimate \eqref{least-squares} and channel estimate in
\eqref{chan_est} to obtain channel estimates of the active users. Note that, we expect to successively obtain better support recovery
performance as we move towards the end of a coherence interval as more CSI becomes available. OMC is concisely formulated in Table I. Before proceeding any further, several remarks are in order.\\

%%%%%%%%%%%%%%%%%%%%%%%%%%%%%%%%%%%%%%%%%%%%%%%%%%%%%%%%%%%%%%%%%%%%%%%%%%%%%%%
\begin{table}[t] \label{table1}\centering
	\begin{tabular}[c]{|l|}
		\hline
		{\bf Table I. OMC Algorithm} \\ \hline
		{\bf Initialization.} At time slot one, form decision statistics $\theta_n$ as in \eqref{dec_stat}, pick the $K$ largest indices as $\hat{\cS}$ and solve the corresponding\\ $\qquad\qquad$reduced-dimension least squares in \eqref{least-squares} and estimate the channels as in \eqref{chan_est}.\\
		{\bf Repeat} for access slots $t=2,3,\ldots,T_s$ \\
		$\qquad$ --$~$Form the decision statistics as in \eqref{otd5}. Select the $K$ decision statistics with largest values and place them in $\cS_1$. \\
		$\qquad$--$~$ Form the decision statistics as in \eqref{dec_stat}. Select the $K$ decision statistics with largest values and place them in $\cS_2$ \\
		$\qquad$--$~$ Set $\hat{\cS}=\cS_1\cup\cS_2$, then run the reduced-dimension least-squares in \eqref{least-squares}. Finally, Form $\|\hat{\bX}^n\|_2^2$ for $n\in\hat{\cS}$. \\
		$\qquad$--$~$ After Picking the largest $K$ values in the previous line, remove the other indices from $\hat{\cS}$ and return $\hat{\cS}$ as the true support.\\
		$\qquad$--$~$ Estimate the channels for devices in $\hat{\cS}$ as in \eqref{chan_est}. \\
		$\qquad$--$~$ Average the newly computed channel estimate with older ones for users belonging to $\Lambda$. \\
		$\qquad$--$~$ Add those users that are activated for the first time to $\Lambda$\\
		{\bf End}\\ \hline
	\end{tabular}
\end{table}
%%%%%%%%%%%%%%%%%%%%%%%%%%%%%%%%%%%%%%%%%%%%%%%%%%%%%%%%%%%%%%%%%%%%%%%%%%%%%%%
\noindent{\bf Remark 1.} We assume known powers in OMC. This assumption is reasonable and practical because path-loss varies slowly over time and thus can be easily estimated beforehand.\\
{\bf Remark 2.} OMC is opportunistic in the sense that it does not schedule
users or force them to send data/training/code for channel estimation.
Instead, it relies on the information it obtains on-demand from the active
users which are random and transmit at will. \\
{\bf Remark 3.} OMC operates independently in different coherence intervals because we assume a block-fading channel
model where the channel changes independently across blocks. \\
{\bf Remark 4.} One might suggest that instead of moving channels of users
with known CSI into the sensing matrix, separate them completely from the
random access process and decode them by applying a e.g., zero-forcing (ZF)
or maximum ratio combining (MRC) to the BS measurements. There exist three
limitations to this approach. Firstly, the channel estimates obtained on a
single random access slot may not be very accurate. As we progress across
random access slots and the same user becomes active multiple times, we
improve our channel estimate. This process can not be carried-out if users
with known CSI are separated. Secondly, if we incur an error in support
estimation, it leads to an erroneous channel estimate. While we can recover
from such errors by averaging CSI at different random access slots that a
particular user becomes active, as done by OMC, separating users with known
CSI leads to error propagation. Thirdly, if $N$ is larger than $M$, ZF is not
practical as inverse of $\bH^T\bH$ either does not exist or is on the order
of $M$ which is too complex to compute. Besides, these users transmit only
sporadically and therefor our efforts in implementing ZF can be completely
wasted. On the other hand, MRC is what we are approximately doing
in our proposed OMC with a modification of an added pseudo-random code.\\
{\bf Remark 5.} There exists a notable difference between the model to which
\cite{baron09} applies trivial pursuit (TP) and our model in \eqref{otd4}.
Specifically, for a fixed $M,L,N,K$, \cite{baron09} has $LM$
measurements and $NM$ unknowns. Note that number of unknowns always exceeds
that of equations and therefore LS is not applicable. On the other
hand, for our problem, in the extreme case, where all users' CSI is known and
moved into the sensing matrix, \eqref{otd3} suggests we get $N$
unknowns and $LM$ equations. For $LM > N$, this model can be solved via LS
but note that its complexity will be of order ${\cal O}(N^3)$. OMC, on the other hand, offers a complexity that is linear in both $M,N$.

%%%%%%%%%%%%%%%%%%%%%%%%%%%%%%%%%%%%%%%%%%%%%%%%%%%%%%%%%%%%%%%%%%%%%%%%%%%%%%
\begin{figure}[t]
	\centering
	
	\begin{minipage}{.5\textwidth}
		\centering
		\captionsetup{width=.8\linewidth}
		\includegraphics[width=.9\linewidth]{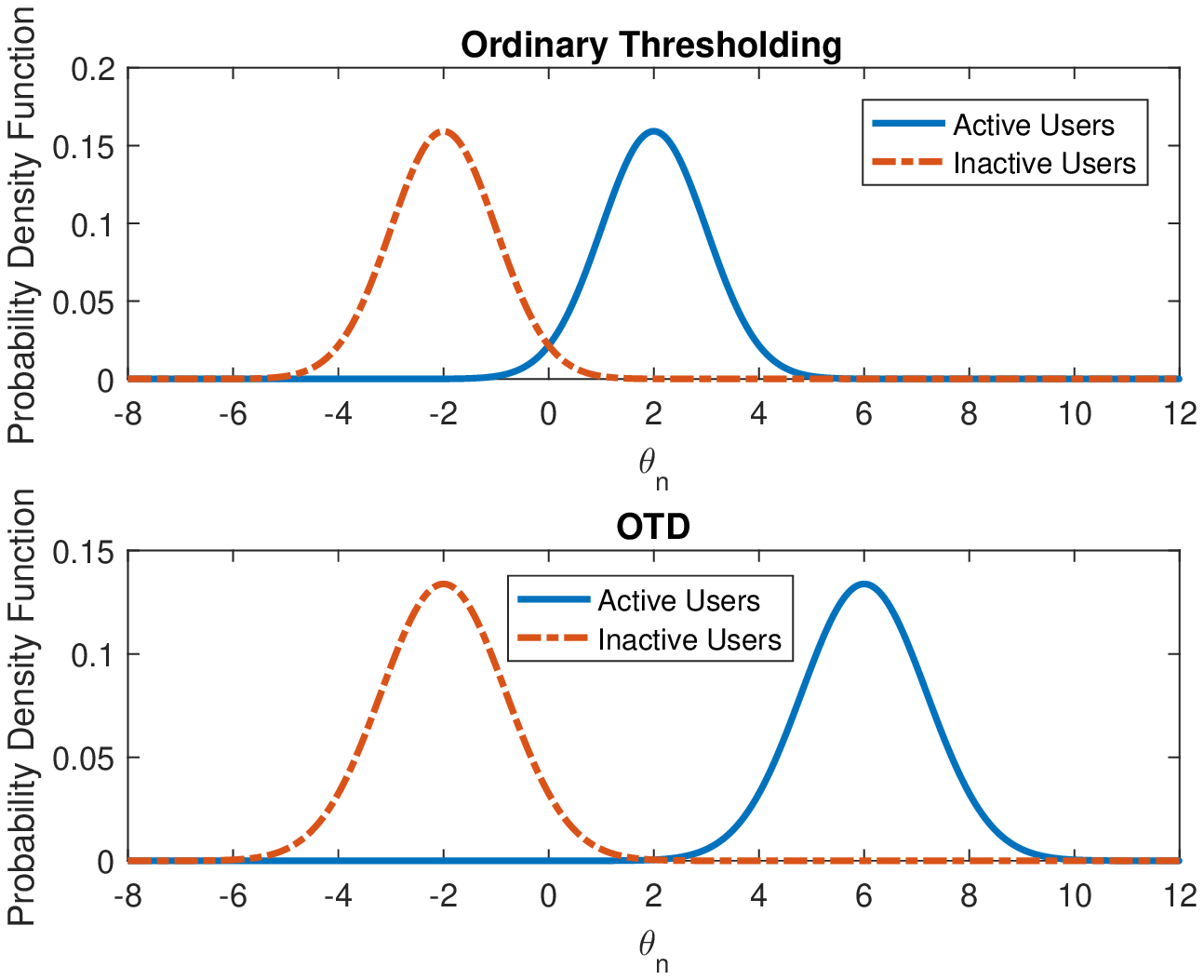}
		\caption{Intuition on why OMC works.}
		\label{fig:fig_3}
	\end{minipage}%
	\begin{minipage}{.5\textwidth}
		\centering
		\captionsetup{width=.8\linewidth}
		\includegraphics[width=.9\linewidth]{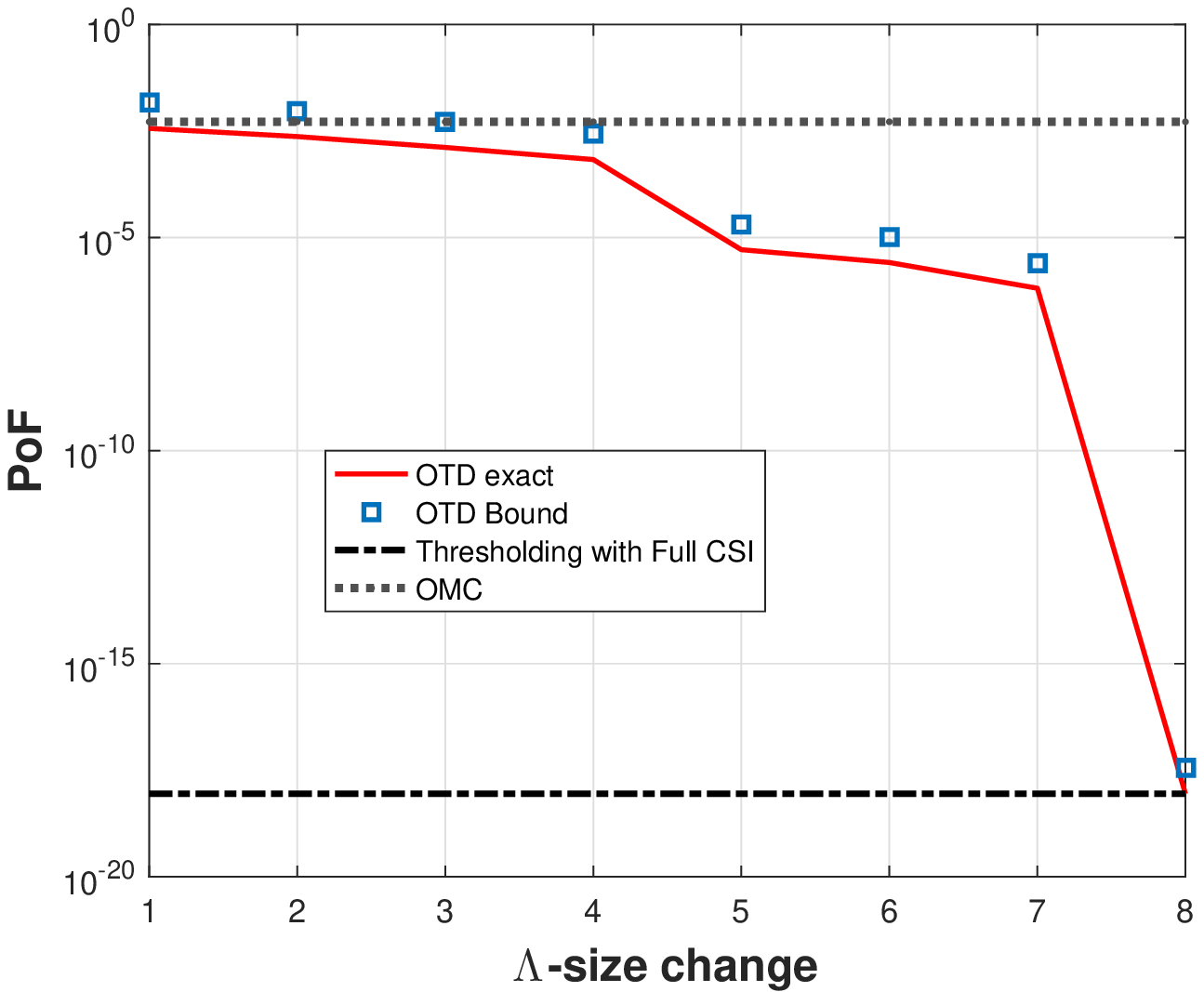}
		\caption{Theoretical evaluation of PoF for OTD and OMC for $L=49510$, $M=50000$, $N=400000$, $K=40$ as a function of $|\Lambda|=[120000,220001, 300000,348000,399600$ $,399800,399950,400000]$.}
		\label{fig_BND_OMC}
	\end{minipage}
\end{figure}
%%%%%%%%%%%%%%%%%%%%%%%%%%%%%%%%%%%%%%%%%%%%%%%%%%%%%%%%%%%%%%%%%%%%%%%%%%%%%%%

Finally, we offer an intuition on why OMC works. OMC performs as illustrated
in Fig. \ref{fig:fig_3}. Initially, when no CSI is available, distribution of
decision statistics $\theta_n$ is plotted in the top plot. Note that the
intersection area in the middle is where errors occur. On the other hand, OMC
moves the mean of $\theta_n$ for active users to a higher value as in the
bottom subplot of Fig. \ref{fig:fig_3}, while slightly increasing the
variance. However, the positive effect of higher mean outdoes the negative
effect of small increase in variance. As a result, performance improves. Note
that this phenomenon occurs only for those users who have been active at
least once in this coherence interval, and decision statistics for inactive
users will remain as in top subplot of Fig. \ref{fig:fig_3}. This figure,
which is verified by analysis that will pursue, suggests that two thresholds
might work better than a single threshold. One separate threshold for users
with no CSI and a second higher threshold for users with CSI. This leads to
opportunistic thresholding detector (OTD) which will be derived in the
following subsection. Finally, an alternative viewpoint on OMC emphasizes its
resemblance to decision feedback equalizer (DFE). In fact, OMC first performs
support recovery. Then, it uses support information to estimate active users
channels. Afterwards, it applies the newly acquired CSI to improve its
support recovery performance in the next random access slot. In a sense, OMC
alternates between support detection and channel estimation which is a
characteristic of DFE-type methods.

\subsection{Opportunistic Thresholding Detector}
One major limitation of OTD compared to OMC is that it demands equal received power from all devices. When powers are not equal, the mean of the decision statistics, as derived in Appendix A, will depend on the powers of active user set, which is unknown to OTD. Hence, suitable thresholds can not be determined. Per Remark 1 in OMC, in IoT networks with low-mobility, power control is achieved with low overhead. Hence, OTD is particularly applicable to such systems.

OTD's initialization step is similar to that of OMC. After the initialization stage, OTD evaluates the decision statistic $\theta_n$ as in \eqref{otd5} then compares them against a threshold. However, two different thresholds are applied to devices with CSI, i.e., those belonging to $\Lambda$, and devices without CSI. If $\theta_n \geq T_{\Lambda}$, the corresponding device with CSI is assumed active while it is assumed inactive otherwise. Similarly, If $\theta_n \geq T_{\bar{\Lambda}}$, where overbar denotes the complement of a set the corresponding device without CSI is assumed active while it is assumed inactive otherwise. The optimum thresholds are derived from performance analysis and are given by
\begin{equation} \label{twothresholds}
T_{\Lambda}=\frac{P\left(3-\frac{1}{L}\right)}{2}+\frac{K}{L}P+\sigma_w^2,\qquad
T_{\bar{\Lambda}}=\frac{P\left(1-\frac{1}{L}\right)}{2}+\frac{K}{L}P+\sigma_w^2
\end{equation}
OTD is concisely formulated in Table II.

%%%%%%%%%%%%%%%%%%%%%%%%%%%%%%%%%%%%%%%%%%%%%%%%%%%%%%%%%%%%%%%%%%%%%%%%%%%%%%%
\begin{table}[t] \label{table1}\centering
	\begin{tabular}[c]{|l|}
		\hline
		{\bf Table II. OTD Algorithm} \\ \hline
		{\bf Initialization.} At time slot one, form decision statistics $\theta_n$ as in \eqref{dec_stat}, pick the $K$ largest indices as $\hat{\cS}$ and solve the corresponding\\ $\qquad\qquad$reduced-dimension least squares in \eqref{least-squares} and estimate the channels as in \eqref{chan_est}.\\
		{\bf Repeat} for access slots $t=2,3,\ldots,T_s$ \\
		$\qquad$ --$~$Form the decision statistics as in \eqref{otd5}. For all $n\in\Lambda$, if $\theta_n \geq T_{\Lambda}$ set device as active and add index to $\cS_1$.\\
		$\qquad$ --$~$ For all $n\in\bar{\Lambda}$, if $\theta_n \geq T_{\bar{\Lambda}}$ set device as active and add index to $\cS_1$. \\
		$\qquad$--$~$ Form the decision statistics as in \eqref{dec_stat}. Select the $K$ decision statistics with largest values and place them in $\cS_2$ \\
		$\qquad$--$~$ Set $\hat{\cS}=\cS_1\cup\cS_2$, then run the reduced-dimension least-squares in \eqref{least-squares}. Form $\|\hat{\bX}^n\|_2^2$ for $n\in\hat{\cS}$ from \eqref{least-squares}. \\
		$\qquad$--$~$ After Picking the largest $K$ values in the previous line, remove the other indices from $\hat{\cS}$ and return $\hat{\cS}$ as the true support.\\
		$\qquad$--$~$ Estimate the channels for devices in $\hat{\cS}$ as in \eqref{chan_est}. \\
		$\qquad$--$~$ Average the newly computed channel estimate with older ones for users belonging to $\Lambda$. \\
		$\qquad$--$~$ Add those users that are activated for the first time to $\Lambda$\\
		{\bf End}\\ \hline
	\end{tabular}
\end{table}
%%%%%%%%%%%%%%%%%%%%%%%%%%%%PERFORMANCE ANALYSIS%%%%%%%%%%%%%%%%%%%%%%%%%%%%%%%
%%%%%%%%%%%%%%%%%%%%%%%%%%%%%%%%%%%%%%%%%%%%%%%%%%%%%%%%%%%%%%%%%%%%%%%%%%%%%%%

\section{Performance Analysis}
While OMC and OTD were designed to outperform existing methods in the small $L$ and large $M$ regime, we are also interested to know if the small $L$ gains carry out to large $L$ values. Here, we perform a rigorous average-case analysis. For the difference between worst-case and average-case analysis please check \cite{eldar10}.

To highlight the novelty of our analysis, we first explain the limitations of current literature. While \cite{gribonval08} used uniform concentration of measure
for Gaussian random variables and Lipschitz functions, this model can not be
applied to our problem because $\bA_m$s as introduced in \eqref{otd4} are not
Gaussian. Furthermore, \cite{gribonval08} focuses on a fixed sensing matrix. As for
\cite{eldar10}, same differences exists. If we now focus on the other model extreme which deals with independent sensing matrices and independent $\bx_m$ entries, one arrives at TP  whose performance was evaluated by a simple application of law of large numbers \cite{baron09}. Unfortunately, our sensing matrices are not independent.

OMC and OTD lie in between the two extremes of fixed and independent sensing matrices. Only in the limit, as $T_s$ grows large, $\bA_m$s become fully uncorrelated. Even uncorrelatedness of sensing matrices will not help because the summands in the decision statistic in \eqref{otd5} are nonlinear functions of sensing matrices entries. Hence, they are neither independent nor uncorrelated.
Consequently, law of large numbers can not be applied. Notably, our analysis of OMC and OTDs will offer, as side results, the performance of thresholding with no CSI and full CSI as they represent special cases of OMC. Proofs for PoF derivation are mainly based on Martingale-theory for concentration of measure \cite[Chapter 2]{wainwright_2019} and are relegated to the appendices. To simplify the derivations, we assume that those users with CSI have a very accurate CSI without noise or any other errors. We drop this assumptions in the simulations and investigate all sources of error such as support detection errors and noise in CSI estimates.

\noindent{\bf Theorem 1 (PoF for OMC)}. For large $M,L,N,K$, PoF for OMC is bounded as
\begin{align}
\textmd{PoF} \leq \label{OMC_UBound}
4\exp \left (\max\left\{\frac{-t}{2b_{nb}}+\ln\left(N-|\Lambda|-K_2\right),\frac{-t}{2b_{ng}}+\ln(K-K_1),\frac{-t}{2b_{fb}}+\ln(|\Lambda|-K_1),\frac{-t_f}{2b_{fg}}+\ln(K_1)\right\}\right)
\end{align}
provided that 
\[\frac{t}{2} > \max\left\{\frac{\nu_{nb}^2}{b_{nb}}, \frac{\nu_{ng}^2}{b_{ng}}, \frac{\nu_{fb}^2}{b_{fb}}\right \}, \qquad \frac{t_f}{2} > \frac{\nu_{fg}^2}{b_{fg}}\]
Here $K_1$ equals the number of active users with CSI, $K_2$ is number of active users without CSI, and other parameters are defined as follows:
\begin{eqnarray*}
t&=&P_{\min}\left(1-\frac{1}{L}\right),\quad t_f=P_{\min}\left(3-\frac{1}{L}\right), \qquad P_{\min}=\underset{k=1,\ldots,K}{\min}~P_k, \quad P_{\max}=\underset{k=1,\ldots,K}{\max}~P_k\\
b_{nb}&=&\max\left\{4\log\log(L),\left(\frac{16\log\log(L)}{M}\right),\left(\frac{4\sigma_w^2L}{MKP_{\max}}\right)\right\}\frac{KP_{\max}}{L} \\
b_{ng}&=&\max\left\{4\log\log(L),\left(\frac{16\log\log(L)}{M}\right),\left(\frac{4\sigma_w^2L}{M(K-1)P_{\max}}\right)\right\}\frac{(K-1)P_{\max}}{L}\\
b_{fb}&=&\max\left\{4\log\log(L),\left(\frac{16\log\log(L)}{M}\right)\right\}\frac{KP_{\max}}{L}\\
b_{fg}&=&\max\left\{4\log\log(L),\left(\frac{16\log\log(L)}{M}\right)\right\}\frac{(K-1)P_{\max}}{L}\\
\nu_{nb}^2&=&\frac{4(L-1)KP_{\max}^2}{L^3}+\frac{64K^2P_{\max}^2(\log\log(L))^2}{ML^2}+\frac{4L\sigma_w^4+32\sigma_w^2KP_{\max}\log\log(L)}{ML}\\
\nu_{ng}^2&=&\frac{4(L-1)(K-1)P_{\max}^2}{L^3}+\frac{64(K-1)^2P_{\max}^2(\log\log(L))^2}{ML^2}\\ & &\qquad+\frac{4L\sigma_w^4+8L\sigma_w^2P_{\max}+32\sigma_w^2KP_{\max}\log\log(L)}{ML}\\ & &\qquad +\frac{3P_{\max}^2+48(K-1)^2P_{\max}^2\frac{(\log\log(L))^2}{L^2}+24(K-1)P_{\max}\frac{(\log\log(L))}{L}}{M}\\
\nu_{fb}^2&=&\frac{4(L-1)KP_{\max}^2}{L^3}+\frac{64K^2P_{\max}^2(\log\log(L))^2}{ML^2}+\frac{4L\sigma_w^4+96\sigma_w^2KP_{\max}\log\log(L)}{ML}\\
& &\qquad +\frac{\sigma_w^4+48K^2P_{\max}^2\frac{(\log\log(L))^2}{L^2}+8\sigma_w^2KP_{\max}\frac{(\log\log(L))}{L}}{M}\\
\nu_{fg}^2&=&\frac{4(L-1)KP_{\max}^2}{L^3}+\frac{64(K-1)^2P_{\max}^2(\log\log(L))^2}{ML^2}+\frac{12\sigma_w^4+8\sigma_w^2\left(15P_{\max}+\frac{12(K-1)P_{\max}\log\log(L)}{L}\right)}{M}\\ & & \hspace{-2cm}+\frac{105 \: P_{\max}^2 + 144 (K-1)^2P_{\max}^2 \frac{\left(\log\log(L)^2\right)}{L^2} + \left(360 \: P_{\max}^2 + 24 \: \sigma_W^2\right)(K-1)P_{\max}  \frac{\left(\log\log(L)\right)}{L}+ 3\sigma_W^4 + 30 \: P_{\max} \sigma_W^2}{M}
\end{eqnarray*}
\emph{Proof:} See Appendix A.

While accurate, the above theorem is hard to interpret. So, we focus on the special case when $M$ grows considerably faster than $L,K,N$.\\
{\bf Corollary 1}. When $M$ increases at a much faster pace than $L,N,K$, PoF for OMC is bounded by
\begin{align}
\textmd{PoF} \leq \label{OMC_UBound_Appx}
4\max\left\{N-|\Lambda|,|\Lambda|\right\}\exp \left (\frac{-P_{\min}(L-1)}{8KP_{\max}\log\log(L)}\right)
\end{align}
Next, we derive the measurement inequality for OMC.\\
{\bf Theorem 2 (Measurement Inequality for OMC)}. For very large $M$, PoF for OMC can be driven below a threshold $\delta$ provided that $L$ is chosen large enough to satisfy the following expression:
\[
 \frac{L-1}{\log\log(L)} \geq \frac{8KP_{\max}}{P_{\min}} \log\left(\frac{4\max\{|\Lambda|,N-|\Lambda|\}}{\delta}\right).
\] 
\emph{Proof:} Equation \eqref{OMC_UBound_Appx} in Corollary 1 is upper bounded by $\delta$ followed by simple algebraic manipulations.

As the first special case of OMC, setting $|\Lambda|=K_1=0$ yields an upper bound on the PoF of ordinary thresholding. Afterwards, a simple measurement inequality for thresholding with no CSI is derived\\
{\bf Corollary 2}. For very large $M$, PoF for ordinary thresholding with no CSI can be made smaller than arbitrary $\delta > 0$ provided that $L$ is chosen larger than a threshold given by:
\[
 \frac{L-1}{\log\log(L)} \geq \frac{8KP_{\max}}{P_{\min}} \log\left(\frac{2(N-K)}{\delta}\right).
\]
A comparison between Theorem 2 and Corollary 2 reveals that measurement inequalities for ordinary thresholding and OMC are from the same order. Thus, in the large $L$ regime, i.e., when measurements are abundant, it is expected that they both perform in a similar fashion. Now, let us derive the measurement inequality for the case of full CSI for all users.\\
{\bf Corollary 3}. For thresholding with full CSI, PoF can be made smaller than arbitrary $\delta > 0$ provided that $L$ is chosen larger than a certain threshold:
\[
\frac{3L-1}{\log\log(L)} \geq \frac{8KP_{\max}}{P_{\min}} \log\left(\frac{2(N-K)}{\delta}\right).
\]
\emph{Proof:} Check Appendix A for the reason behind replacing $t$ with $t_f$. The rest of the proof is easily obtained by replacing $|\Lambda|=N,K_1=K$ in Theorem 1. 

Compared to OMC and thresholding with no CSI, thresholding with full-CSI demands one third of the measurements to obtain the same performance which is a consequence of harvesting diversity in sensing matrices. The conclusions are intriguing, as it seems that OMC does not have a clear gain in the large $L$ regime, while its limiting case of full CSI reduces the needed measurements by a factor of three. Surprisingly though, OMC performs superior to ordinary thresholding as well as state of the art algorithms in the low $L$ regime as witnessed by our numerical results. Now, let us focus on OTD.\\
{\bf Theorem 3 (PoF for OTD)}. For large $M,L,N,K$, PoF for OTD is bounded as
\begin{align}
\textmd{PoF} \leq \label{OTD_UBound}
4\exp \left (\max\left\{\frac{-t}{2b_{nb}}+\ln\left(N-|\Lambda|-K_2\right),\frac{-t}{2b_{ng}}+\ln(K-K_1),\frac{-t_f}{2b_{fb}}+\ln(|\Lambda|-K_1),\frac{-t_f}{2b_{fg}}+\ln(K_1)\right\}\right)
\end{align}
provided that 
\[\frac{t}{2} > \max\left\{\frac{\nu_{nb}^2}{b_{nb}}, \frac{\nu_{ng}^2}{b_{ng}}\right\}, \qquad \frac{t_f}{2} > \max\left\{ \frac{\nu_{fg}^2}{b_{fg}},\frac{\nu_{fb}^2}{b_{fb}}\right\}.\]

\emph{Proof:} Check Appendix B.\\
Note that OTD operates in the equal power mode only, i.e., when power control is applied. Therefore, for OTD $P_{\min}=P_{\max}=P$. Again, the expression for PoF is very complicated. To simplify, we let $M$ grow very large and approximate the PoF as
\begin{align}
\textmd{PoF} \leq \label{OTD_UBound_Appx}
4\max\left\{(N-|\Lambda|)\exp \left (\frac{-(L-1)}{8K\log\log(L)}\right),|\Lambda|\exp \left (\frac{-(3L-1)}{8K\log\log(L)}\right)\right\}
\end{align}
{\bf Theorem 4 (Measurement Inequality for OTD)}. For very large $M$, PoF for OTD can be driven below a threshold $\delta$ provided that $L$ is chosen large enough to satisfy the following expressions.\\
When $|\Lambda|$ is small,
\[
\frac{L-1}{\log\log(L)} \geq 8K \log\left(\frac{4 (N-|\Lambda|)}{\delta}\right).
\]
When $|\Lambda|$ is large
\[
\frac{3L-1}{\log\log(L)} \geq 8K \log\left(\frac{4 (|\Lambda|)}{\delta}\right).
\]
A comparison with previous results reveals that OTD performs similar to ordinary thresholding when number of users with CSI is small. On the other hand, its performance grows close to thresholding with full CSI when number of users with CSI increases. This conclusion is in contrast to OMC whose large regime performance is almost the same as ordinary thresholding with no CSI. These observations are summarized in Fig. \ref{fig_BND_OMC} where analytical PoF curves for OTD and OMC are plotted. While OTD is superior to OMC in this regard, it has two major limitations. First, it requires power control. Secondly, as corroborated by the simulations, it requires a larger $M$ to perform satisfactorily compared to OMC. 

\section{Numerical Results}
We should point out that the chief significance of OMC / OTD lies in their superior performance when measurement size, $L$, is small. The small $L$ regime is difficult to analyze mathematically. Therefore, we complement the previous section with extensive numerical simulations and compare OMC / OTD versus state of the art algorithms. Towards this goal, we have selected M-SBL, G-OMP, and two versions of AMP as alternatives. First AMP version is taken from \cite{baron11} which offers an algorithm particularly suited to the large $M$ values and is referred to as AMP 1. The second AMP version is borrowed from \cite{Chen18} and is referred to as parallel AMP. We refer to it as AMP 2. We did not compare with CoSaMP as it needs $L \geq 3K$ which is too large an $L$. In addition, we did not consider AMP with vector denoiser proposed by \cite{Chen18}, as it demands $M$ to assume small values. Otherwise, the update for $\tau$ becomes numerically unstable. 

\subsection{Performance versus Random Access Slot}
\begin{figure}[t]
	\centering
	
	\begin{minipage}{.5\textwidth}
		\centering
		\captionsetup{width=.8\linewidth}
		\includegraphics[width=.9\linewidth]{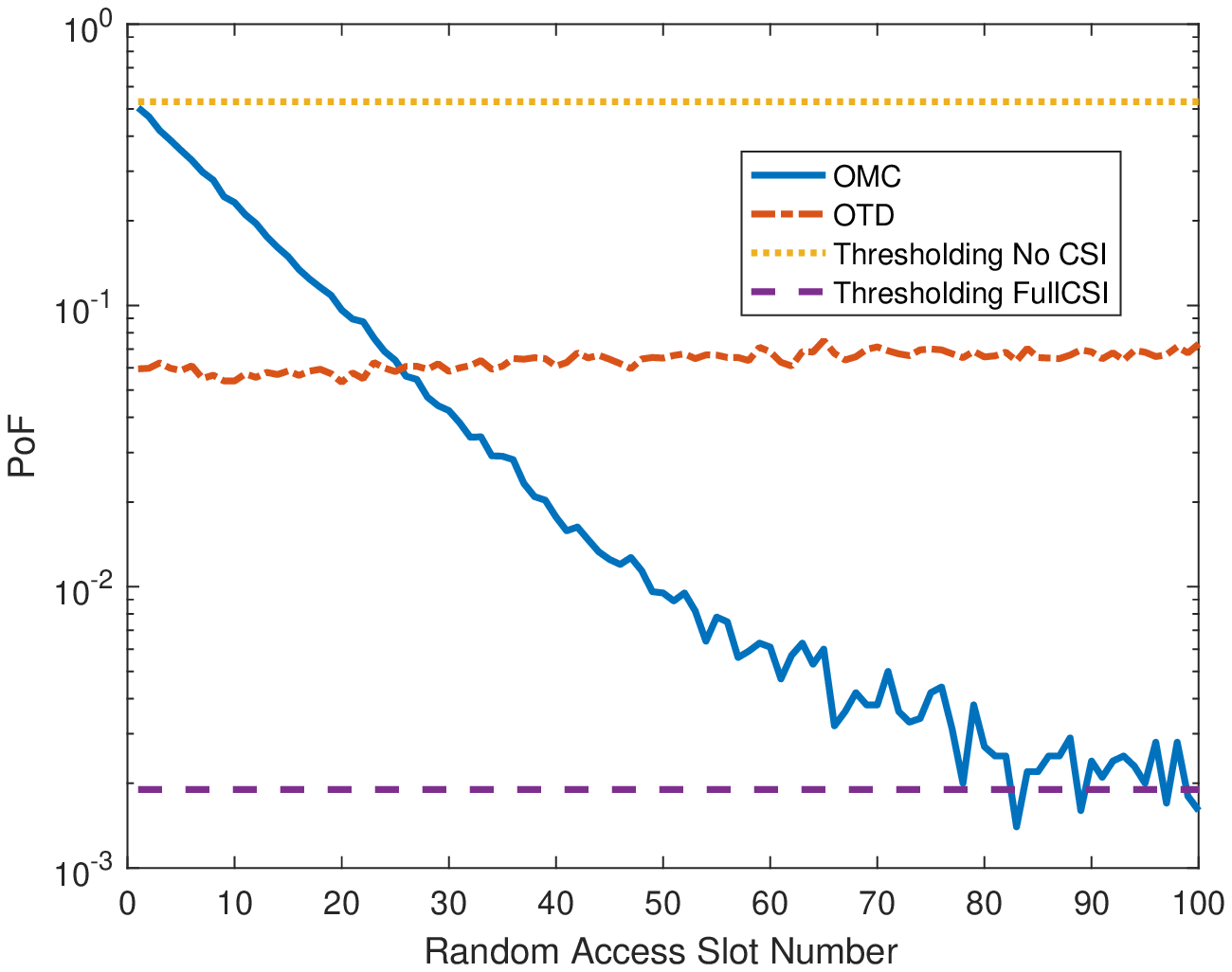}
		\caption{OMC, OTD performance versus their limiting algorithms for $L=35$, $M=256$.}
		\label{fig_1_256}
	\end{minipage}%
	\begin{minipage}{.5\textwidth}
		\centering
		\captionsetup{width=.8\linewidth}
		\includegraphics[width=.9\linewidth]{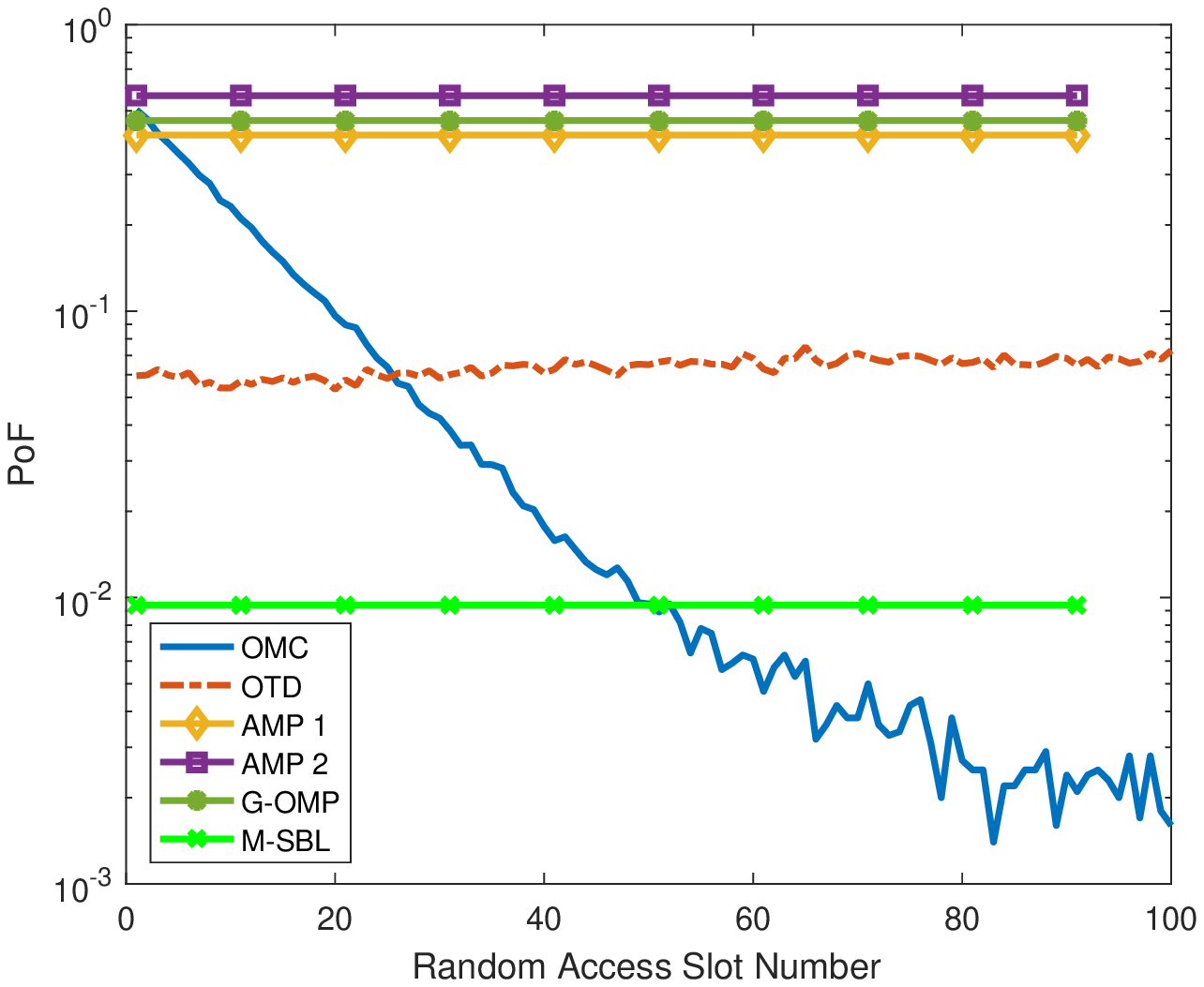}
		\caption{OMC, OTD performance versus state of the art for $L=35$, $M=256$.}
		\label{fig_2_256}
	\end{minipage}
\end{figure}

We selected $N=200,~K=20,~L=35,~M=256$ for our first numerical experiment. Code matrix $\bC$ entries were chosen IID Rademacher which assumed $\pm\{1/\sqrt{L}\}\}$. MIMO channels were selected as IID Gaussian with mean zero and variance $\sqrt{P_k}$. An equal power setup is considered first where the SNR was selected as
\[
\frac{P_k\|\bc_k\|_2^2}{E[\|\bw_m\|_2^2]}=\frac{P_k}{L\sigma_w^2}=P=1
\]
equivalent to 0 dB. Fig. \ref{fig_1_256} plots PoF versus random access slot for OMC, OTD, Ordinary Thresholding, and Thresholding with full CSI. As can be seen from the figure, Thresholding with full CSI which exploits diversity in sensing matrices greatly outperforms ordinary thresholding. OMC and OTD fall in between the two extremes. Initially, OMC performs similar to ordinary thresholding but as time advances and more channels are estimated, its performance improves and converges to that of thresholding with full CSI. Unlike OMC, OTD performance is almost fixed across random access slots and it seems that it does not benefit from diversity in sensing matrices. We illustrate that this is not the case in later figures. Fig. \ref{fig_2_256} compares OMC / OTD versus state of the art. This figure reveals that $L$ is selected too small for G-OMP, AMP 1,and AMP 2. It is only M-SBL that performs satisfactorily. However, provided enough time slots pass and enough CSI is collected, OMC can outperform M-SBL. It should be noted that M-SBL, AMP 1, and AMP 2 are iterative and we performed 10 iterations of each, while for AMP 2 we also run 10 iterations for the outer messages which amounts to 100 iterations overall. Two limitations of our competitors should be pointed out. First one is their increased complexity. While OTD and OMC are one-shot thresholding algorithms that are greedy and do not require iterations, M-SBL, AMP 1, and AMP 2 do need many iterations to converge. In order to make a fair comparison complexity-wise, we limited the number of iterations to 10. Second issue is that none of these algorithms exploit diversity in sensing matrices and hence their performance do not improve across random access slots.

%%%%%%%%%%%%%%%%%%%%%%%%%%%%%%%%%%%%%%%%%%%%%%%%%%%%%%%%%%%%%%%%%%%%%%%%%%%%%%
%%%%%%%%%%%%%%%%%%%%%%%%%%%%%%%%%%%%%%%%%%%%%%%%%%%%%%%%%%%%%%%%%%%%%%%%%%%%%%%

\begin{figure}[t]
	\centering
	
	\begin{minipage}{.5\textwidth}
		\centering
		\captionsetup{width=.8\linewidth}
		\includegraphics[width=.9\linewidth]{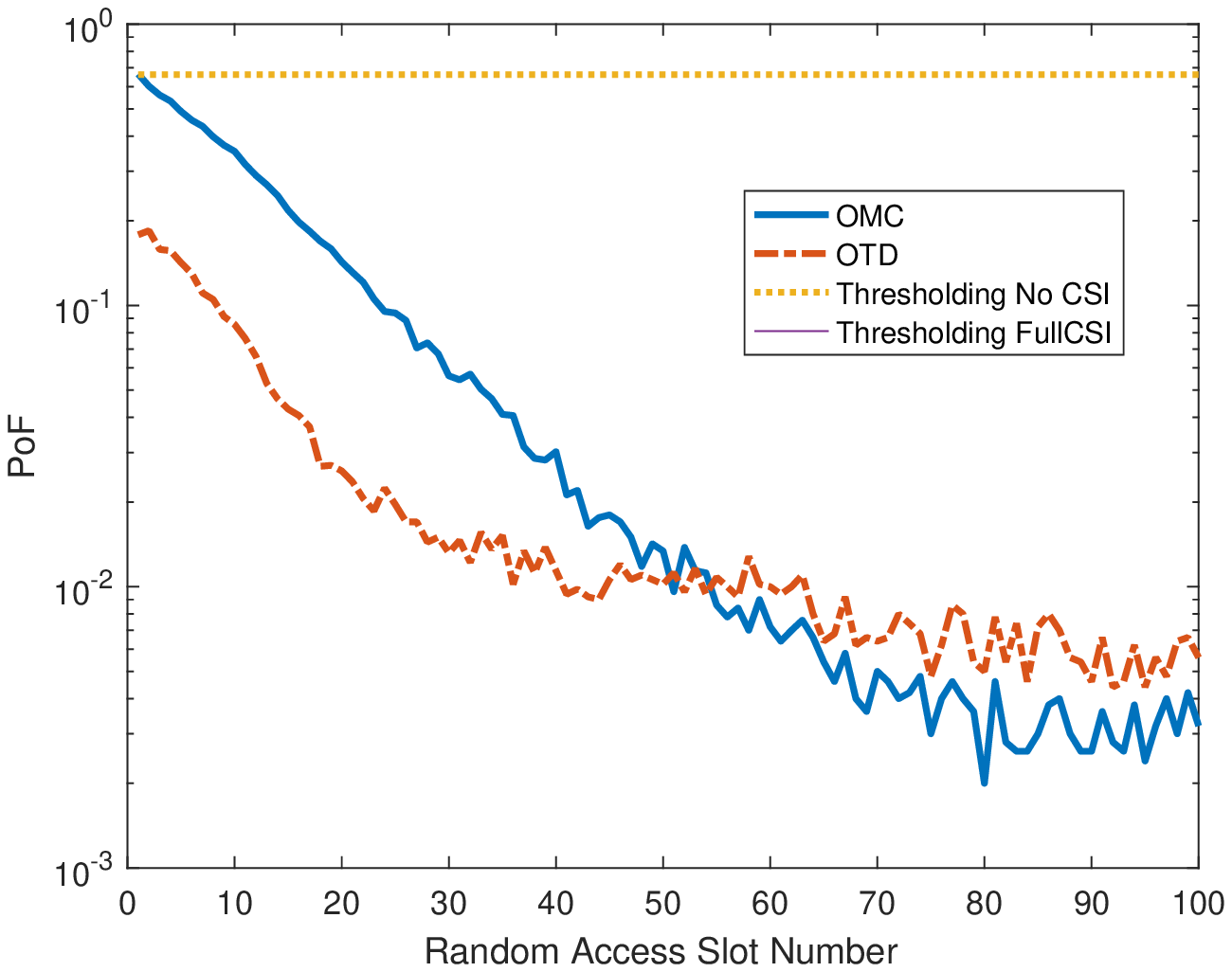}
		\caption{OMC, OTD performance versus their limiting algorithms for $L=30$, $M=1024$.}
		\label{fig_1_1024}
	\end{minipage}%
	\begin{minipage}{.5\textwidth}
		\centering
		\captionsetup{width=.8\linewidth}
		\includegraphics[width=.9\linewidth]{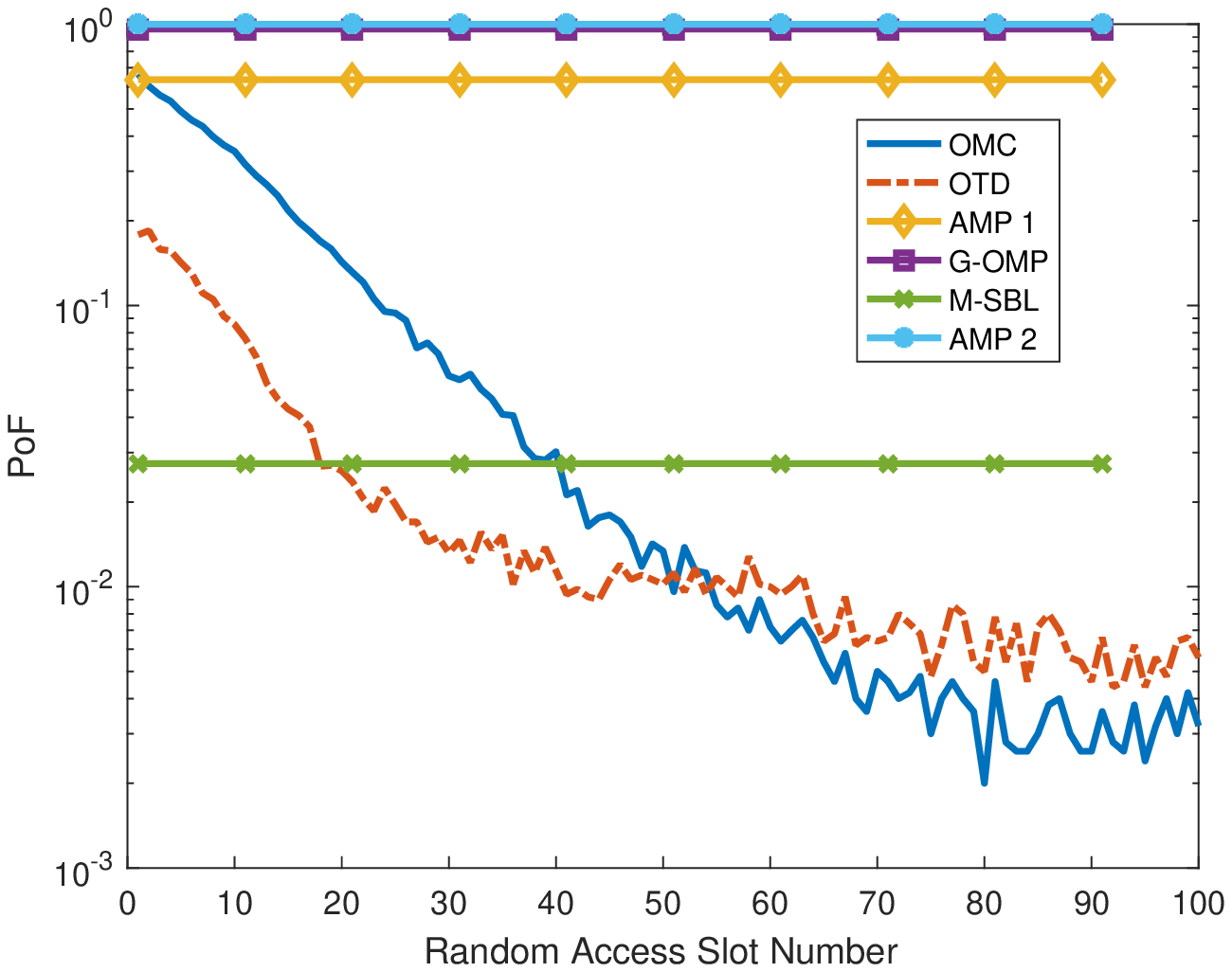}
		\caption{OMC, OTD performance versus state of the art for $L=30$, $M=1024$.}
		\label{fig_2_1024}
	\end{minipage}
\end{figure}

In Figs. \ref{fig_1_1024}, \ref{fig_2_1024} we plot the same algorithms as in previous figures but with $M=1024$ which is a four-fold increase. At the same time, $L$ is reduced to $30$. Fig. \ref{fig_1_1024} offers two significant differences versus Fig. \ref{fig_1_256}. First, Thresholding with full CSI yields a probability of error equal to zero in 10000 Monte Carlo runs, which equals a resolution of $10^{-4}$, hence it is not plotted. Secondly, OTD now improves as more CSI is collected similar to OMC. Note that both of these algorithms somehow rely on law of large numbers and therefore require $M$ to be sufficiently large. Due to the different nature of the two algorithms the improvement due to diversity happens at a smaller $M$ for OMC compared to OTD. Fig. \ref{fig_2_1024} again illustrates the superior performance of OMC / OTD versus state of the art.

%%%%%%%%%%%%%%%%%%%%%%%%%%%%%%%%%%%%%%%%%%%%%%%%%%%%%%%%%%%%%%%%%%%%%%%%%%%%%%
%%%%%%%%%%%%%%%%%%%%%%%%%%%%%%%%%%%%%%%%%%%%%%%%%%%%%%%%%%%%%%%%%%%%%%%%%%%%%%%

Finally, non-equal power scenario is considered were the $N$ devices uniformly assume values for power in a 6 dB power spread. Note that OTD can not operate with non-equal powers. Hence, we plot OMC versus several alternatives in Fig. \ref{fig_nonequal}. Here, we have assumed $M=4096, L=30$, and plotted $T_s=400$ random access slots versus $T_s=100$ we had before. Note that OMC converges very slowly compared to previous figures and even after 400 random access slots, there remains a significant gap between OMC and the thresholding with full CSI. Still, OMC performs better than G-OMP and M-SBL provided enough random access slots have passed.

\subsection{Performance versus changes in $M,L$}
\begin{figure}[t]
	\centering
	
	\begin{minipage}{.5\textwidth}
		\centering
		\captionsetup{width=.8\linewidth}
		\includegraphics[width=.9\linewidth]{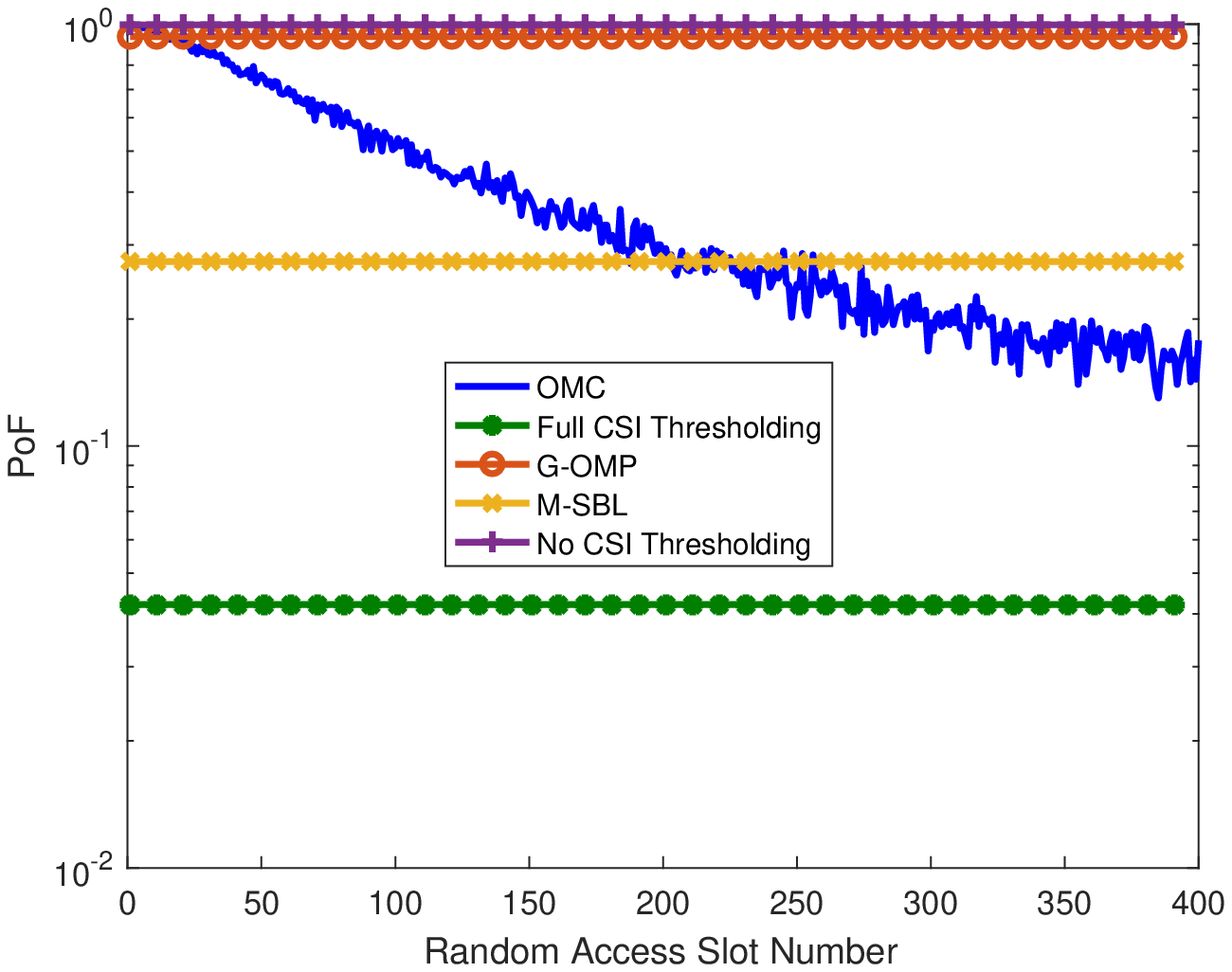}
		\caption{OMC performance versus state of the art for $L=30$, $M=4096$ and power spread of 6 dB.}
		\label{fig_nonequal}
	\end{minipage}%
	\begin{minipage}{.5\textwidth}
		\centering
		\captionsetup{width=.8\linewidth}
		\includegraphics[width=.9\linewidth]{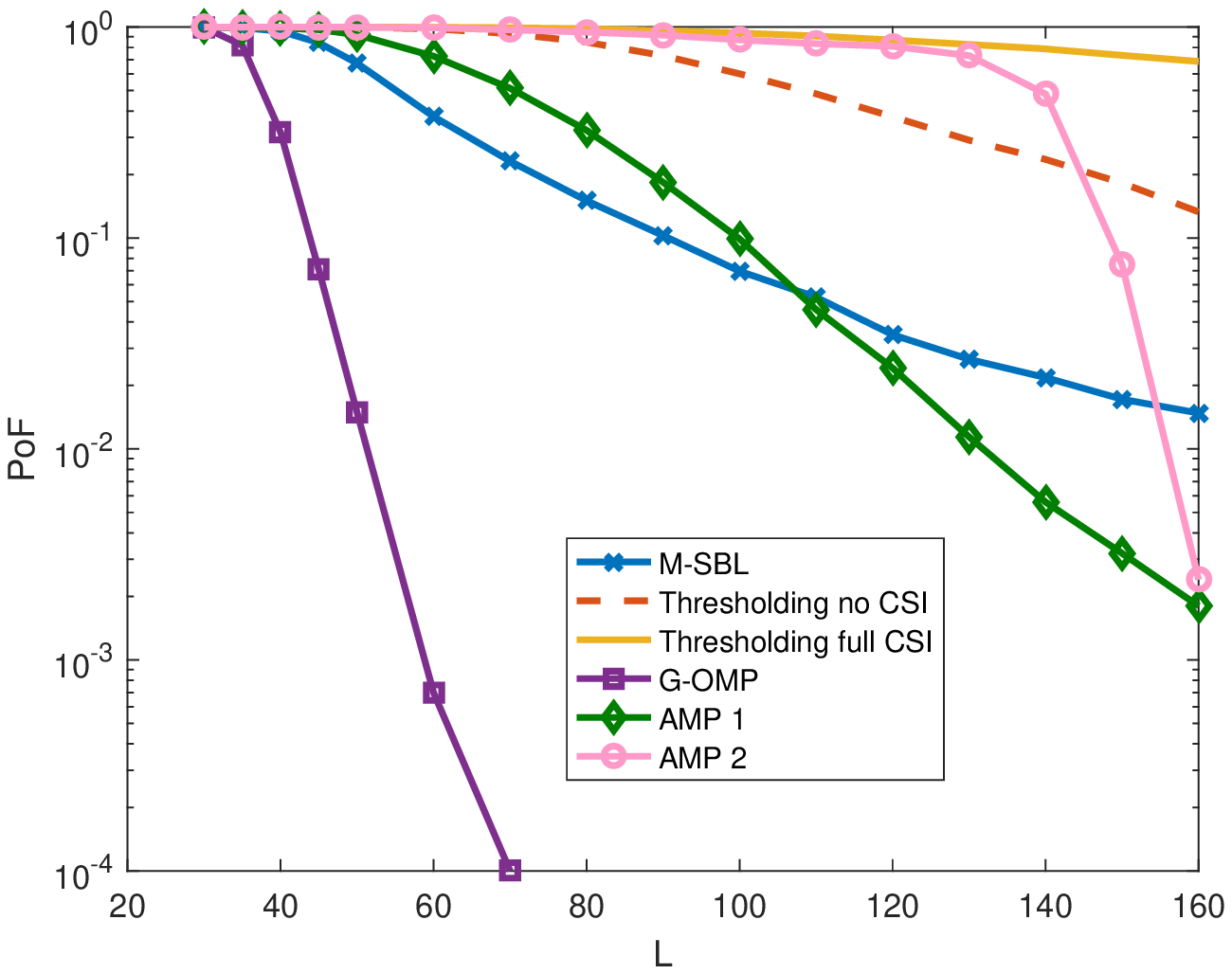}
		\caption{Performance of various algorithms for $M=16$ as a function of $L$}
		\label{fig_L1_16}
	\end{minipage}
\end{figure}

To provide general intuition on how all these competing algorithms fair against one another, we have plotted performance for a range of values of $M,L$. In the first experiment, we have fixed $M=16$ and let $L$ grow from 30 to 160. Results are plotted in Fig. \ref{fig_L1_16}. Note that G-OMP, M-SBL, and AMP 1 offer the best performance. Curiously, thresholding with no CSI performs better than thresholding with full CSI. This phenomenon can be attributed to the fact that by using CSI, we are increasing the decision statistics variance. This negative effect should be compensated by diversity in sensing matrices. That is, as $M$ grows large, measure will concentrate around the mean and because CSI shifts the mean of the decision statistics for active users to a higher value, we get a smaller PoF. However, $M=16$ is not large enough to ensure concentration occurs and thus decision statistics can assume values that are far from mean due to large variance. Therefore, the larger variance leads to a poorer probability of error. The intuition recommends that when $M$ is small existing alternatives will outperform the OMC / OTD.

%%%%%%%%%%%%%%%%%%%%%%%%%%%%%%%%%%%%%%%%%%%%%%%%%%%%%%%%%%%%%%%%%%%%%%%%%%%%%%
%%%%%%%%%%%%%%%%%%%%%%%%%%%%%%%%%%%%%%%%%%%%%%%%%%%%%%%%%%%%%%%%%%%%%%%%%%%%%%%

\begin{figure}[t]
	\centering
	
	\begin{minipage}{.5\textwidth}
		\centering
		\captionsetup{width=.8\linewidth}
		\includegraphics[width=.9\linewidth]{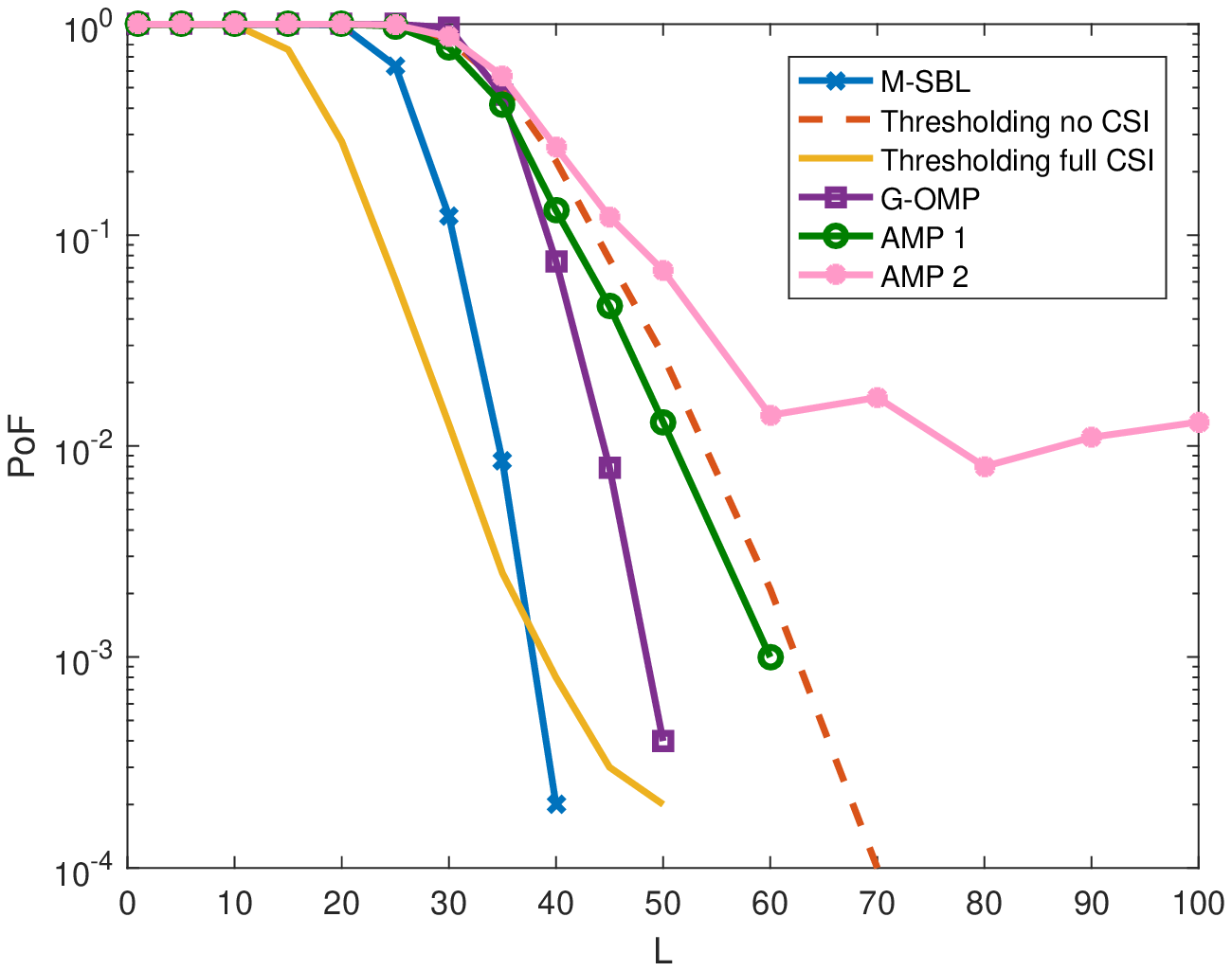}
		\caption{Performance of various algorithms for $M=256$ as a function of $L$.}
		\label{fig_L2_256}
	\end{minipage}%
	\begin{minipage}{.5\textwidth}
		\centering
		\captionsetup{width=.8\linewidth}
		\includegraphics[width=.9\linewidth]{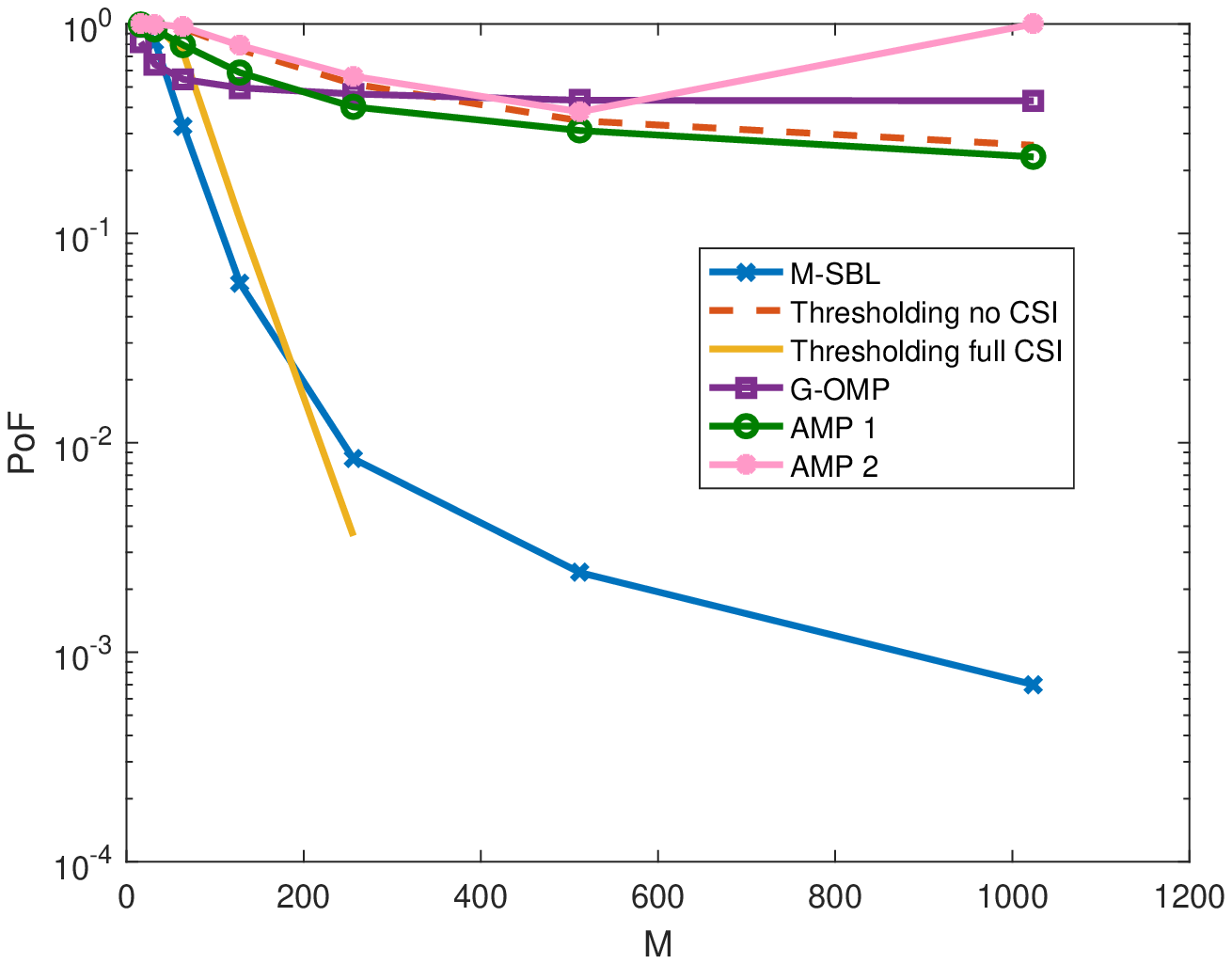}
		\caption{Performance of various algorithms for $L=35$ as a function of $M$}
		\label{fig_M1_L35}
	\end{minipage}
\end{figure}

In the second experiment, we set $M=256$ and plot probability of error in support recovery versus $L$ values. Fig. \ref{fig_L2_256} plots the results of this experiment. This figure suggests that $M=256$ is large enough to offer needed diversity so that thresholding with full CSI outperforms the other algorithms. We have not plotted the OTD / OMC here as their performance depends on the number of random access slot we are in. However, as demonstrated in the previous subsection, their performance begins with ordinary thresholding and converges to that of thresholding with full CSI. This figure shows the merits that diversity in sensing matrices provides.

Finally, we fix $L=35$ and let $M$ increase. Results are plotted in Fig. \ref{fig_M1_L35}. It can be seen that thresholding with full CSI outperforms all the other methods for large $M$ values. M-SBL performs the best in small $M$. The bounce back of AMP 2 for large $M$ is attributed to numerical unstability of AMP 2 for large $M$. Note that in AMP 2, 1023 probabilities should be multiplied and then normalized to one, and this product might well go below the resolution of MATLAB leading to 0 over 0 and NaN components.

%%%%%%%%%%%%%%%%%%%%%%%%%%%%%%%%%%%%%%%%%%%%%%%%%%%%%%%%%%%%%%%%%%%%%%%%%%%%%%
%%%%%%%%%%%%%%%%%%%%%%%%%%%%%%%%%%%%%%%%%%%%%%%%%%%%%%%%%%%%%%%%%%%%%%%%%%%%%%%

To conclude this section, we observe that for small $L$ and large $M$ regime, diversity in sensing matrices, collected by OMC, OTD, and Thresholding with full CSI, which determines the performance limit, can have a huge impact on performance of MMV support recovery algorithms when probability of correct support recovery is the figure of merit.

\section{Conclusion}
An integration of on-off random access channel with massive MIMO was
investigated, where it is known that active users detection amounts to
support recovery for a multiple measurement vector (MMV) problem with a fixed
sensing matrix and common sparse support. Motivated by TP, which collects diversity in sensing matrices, we first offered a recommendation on ways to de-correlate sensing matrices. Afterwards, we introduced two novel thresholding detectors, namely OMC and OTD, that can collect the diversity from partially de-correlated sensing matrices. The proposed algorithms were mathematically analyzed and upper bounds on their PoF were derived along with relevant measurement inequalities. It was revealed that OTD carries its gains to the large $L$ regime, while OMC loses its edge. On the other hand, OTD requires power control, while OMC can handle inequal powers. Finally, extensive simulations corroborated the superior performance of both OMC and OTD for the small $L$ and large $M$ regime compared to the state of the art.

\section*{Appendices}
\section*{Appendix A. Derivation of PoF for OMC}
The PoF is in general defined as \cite{rauhut2008compressed}
\begin{eqnarray}\label{PoF}
\textrm{PoF} & = & p\left(\min_{n\in \mathcal{S}}\:\theta_n \leq \max_{n\in  \bar{\mathcal{S}} } \:\theta_n\right) \leq p\left(\min_{n\in  \mathcal{S}}\:\theta_n \leq \rho_1\right) + p\left(\max_{n\in \bar{\mathcal{S}} } \:\theta_n \geq \rho_2\right) \nonumber \qquad ; {\rho_1\geq\rho_2\geq0} \\ & \leq &p\left(\bigcup_{n\in \mathcal{S}}\:\left\{\theta_n \leq \rho_1\right\}\right) + p\left(\bigcup_{n\in \bar{\mathcal{S}} } \:\left\{\theta_n \geq \rho_2\right\}\right) \leq\sum_{n\in \mathcal{S}}\:p\left(\theta_n \leq \rho_1\right) + \sum_{n\in \bar{\mathcal{S}}} \:p\left(\theta_n \geq \rho_2\right)\nonumber \\ & \leq & \sum_{n\in \bar{\mathcal{S}}\cap\bar{\Lambda}} \:p\left(\theta_n \geq \rho_2\right) + \sum_{n\in \mathcal{S}\cap \bar{\Lambda}}\:p\left(\theta_n \leq \rho_1\right) + \sum_{n\in \bar{\mathcal{S}}\cap\Lambda} \:p\left(\theta_n \geq \rho_2\right) + \sum_{n\in \mathcal{S}\cap \Lambda}\:p\left(\theta_n \leq \rho_1\right) 
\end{eqnarray}
where, $\theta_n$ is test statistics defined in \eqref{otd5}. In the second line, we have used union bound, while the first line benefited from the following Lemma whose proof is omitted.\\
{\bf Lemma 1.} For any two positive random variables $X,Y$, we have $p(X \leq Y)\leq p(X\leq \rho_1)+p(Y\geq \rho_2)$ where $\rho_1 \geq \rho_2 \geq 0$ are arbitrary constants.\\
In the rest of this appendix, we bound the terms in \eqref{PoF}. Note that the four terms in \eqref{PoF} correspond to respectively inactive users without CSI, active users without CSI, inactive users with CSI, and active users with CSI. Let us derive the mean of decision statistics for each of these four terms first. 
Note that we have $\by_m=\sum_{k\in\cS}\sqrt{P_k}H_{m,k}\bc_k+\bw_m$. Upon replacing $\by_m$ from above into the decision statistics in \eqref{otd5} and taking expected values, we arrive at the following equations. For inactive users with no CSI:
\[
{\rm E}[\theta_n]={\rm E}\left[\frac{1}{M} \sum_{m=1}^{M} \left(\bc_n^T\by_m\right)^2\right]
=\frac{\sum_{k\in\cS}P_k}{L}+\sigma_w^2.
\]
For active users with no CSI:
\[
{\rm E}[\theta_n]={\rm E}\left[\frac{1}{M} \sum_{m=1}^{M} \left(\bc_n^T\by_m\right)^2\right]
=P_n\left(1-\frac{1}{L}\right)+\frac{\sum_{k\in\cS}P_k}{L}+\sigma_w^2.
\]
For inactive users with CSI:
\[
{\rm E}[\theta_n]={\rm E}\left[\frac{1}{M} \sum_{m=1}^{M} \left(H_{m,n}\bc_n^T\by_m\right)^2\right]
=\frac{\sum_{k\in\cS}P_k}{L}+\sigma_w^2.
\]
Finally, for active users with CSI
\[
{\rm E}[\theta_n]={\rm E}\left[\frac{1}{M} \sum_{m=1}^{M} \left(H_{m,n}\bc_n^T\by_m\right)^2\right]
=P_n\left(3-\frac{1}{L}\right)+\frac{\sum_{k\in\cS}P_k}{L}+\sigma_w^2.
\]
Note that mean of the active users is higher than nonactive users, while this gap is larger for users with CSI compared to those without CSI. If concentration of measure occurs for OMC, measure should concentrate around mean of $\theta_n$. This guides us on how to choose $\rho_1,\rho_2$. To ensure concentration of measure phenomenon is captured in all four cases, we select
\[
\rho_1:=P_{\min}\left(1-\frac{1}{L}\right)+\frac{\sum_{k\in\cS}P_k}{L}+\sigma_w^2-
\frac{t}{2}, \qquad\qquad \rho_2=\frac{\sum_{k\in\cS}P_k}{L}+\sigma_w^2+\frac{t}{2}.\] 
and $t:=P_{\min}(1-1/L)$. Next, let us begin by upper bounding the first term in \eqref{PoF}. 

\subsection{Inactive users without CSI}
First term in \eqref{PoF}, which belongs to inactive users without CSI, is re-written as follows:
\begin{equation} \label{prob_nocsi}
p\left(\theta_n \geq \rho_2\right)=p\left(\theta_n-{\rm E}[\theta_n]\geq\rho_2-{\rm E}[\theta_n]\right)=p\left(\theta_n-{\rm E}[\theta_n]\geq \frac{t}{2}\right)
\end{equation}
To bound \eqref{prob_nocsi}, we define the following  Martingale difference sequences (MDS):
\[
\theta_n-{\rm E}[\theta_n]=\left(\theta_n-{\rm E}[\theta_n|\bC,\bH]\right)+\left({\rm E}[\theta_n|\bC,\bH]-{\rm E}[\theta_n|\bC]\right)+\left({\rm E}[\theta_n|\bC]-{\rm E}[\theta_n]\right):=D_3+D_2+D_1
\]
Next, we show that $D_1,D_2,D_3$ are sub-exponential and derive their corresponding parameters which are $\nu_1^2,\nu_2^2,\nu_3^2$ and $b_1,b_2,b_3$ and then apply Theorem 2.19 in \cite{wainwright_2019}. We begin with $D_1$,
\begin{align}\label{d_1}
D_1 & = \textmd{E}\left[\theta_n|\bC\right]-\textmd{E}\left[\theta_n\right] = \displaystyle \sum_{\substack{k\in \cS}} P_k \left(\bc_k^T\bc_n\right)^2 - \frac{1}{L}\sum_{\substack{k\in \cS}} \: P_k
\end{align}
Upon defining $S_{L}:=\bc_k^T\bc_n=\sum_{l=1}^{L} c_k(l)c_n(l)$, we observe that $LS_L$ is a symmetric random walk of length $L$ with unit steps which have unit variance. We invoke the law of iterated logarithm which states that
\begin{equation} \label{random_walk}
0\leq S_L^2=(\bc_k^T\bc_n)^2\leq \frac{2(1+\epsilon)^2\log\log(L)}{L}\leq\frac{4\log\log(L)}{L},\qquad (a.s.)
\end{equation}
where the last inequality is obtained by selecting $(1+\epsilon)^2\leq 2$. Thus, we have shown that $D_1$ is bounded almost surely, which means that Bernstein-type bounds can be applied to $D_1$. Using the Bernstein bound in \cite[pp. 27,28]{wainwright_2019}, we deduce that $D_1 \sim \text{sub-exponential}\left(\nu_{1},b_1\right)= \text{sub-exponential}\left(\sqrt{2}\sigma_{1},2b\right)$ where
\begin{align}\label{varD1BUCGM}
\sigma_{1}^2  = \textmd{E}[D_1^2]= \frac{2(L-1)}{L^3} \left(\displaystyle \sum_{\substack{k\in \cS}} P_k^2\right), \qquad \nu_{1}^2 = 2\sigma_{1}^2 = \frac{4(L-1)}{L^3} \left(\displaystyle \sum_{\substack{k\in \cS}} P_k^2\right)\leq \frac{4KP_{\max}^2}{L^2}
\end{align}
Also, parameter \textit{b} is determined as the maximum of absolute lower and upper bounds of $D_1$ \cite{boucheron2013concentration}, which yields (c.f. \eqref{d_1} and \eqref{random_walk}):
\begin{align}\label{bD1BUCGM}
b_1 & = 2 \Big(\frac{4\log\log(L)}{L}-\frac{1}{L}\Big)\displaystyle \sum_{\substack{k\in \cS}} P_k \leq \frac{8KP_{\max}\log\log(L)}{L}
\end{align}

After some algebraic manipulation, the second term of MDS is written as
\begin{align}\label{UCGMBD2}
D_2 & = \frac{1}{M}\sum_{m=1}^M\: \underbrace{\left[\textbf{z}_m^T  \textbf{Q}\: \textbf{z}_m - \mbox{Trace(\textbf{Q})}\right]}_{\xi_{m,n}}
\end{align}
where we have assumed that $\cS:=\{k_1,k_2,\ldots,k_K\}$, and $\textbf{z}_m$, and $\textbf{Q}$ are defined as follows
\begin{align*}
\textbf{Q} &:=
\begin{bmatrix}
P_{k_1} \left(\bc_{k_1}^T\bc_n\right)^2 & \sqrt{P_{k_1}}\sqrt{P_{k_2}} \left(\bc_{k_1}^T\bc_n\right)\left(\bc_{k_2}^T\bc_n\right) & \ldots
& \sqrt{P_{k_1}}\sqrt{P_{k_K}} \left(\bc_{k_1}^T\bc_n\right)\left(\bc_{k_K}^T\bc_n\right) \\
\sqrt{P_{k_2}}\sqrt{P_{k_1}} \left(\bc_{k_2}^T\bc_n\right)\left(\bc_{k_1}^T\bc_n\right) & P_{k_2} \left(\bc_{k_2}^T\bc_n\right)^2 & \ldots
& \sqrt{P_{k_2}}\sqrt{P_{k_K}} \left(\bc_{k_2}^T\bc_n\right)\left(\bc_{k_K}^T\bc_n\right) \\
\vdots & \vdots & \ddots
& \vdots \\
\sqrt{P_{k_K}}\sqrt{P_{k_1}} \left(\bc_{k_K}^T\bc_n\right)\left(\bc_{k_1}^T\bc_n\right) & \sqrt{P_{k_K}}\sqrt{P_{k_2}} \left(\bc_{k_K}^T\bc_n\right)\left(\bc_{k_2}^T\bc_n\right) & \ldots
& P_{k_K} \left(\bc_{k_K}^T\bc_n\right)^2
\end{bmatrix}\\ 
\textbf{z}_m &:=\left[\begin{array}{cccc} H_{m,{k_1}} & H_{m,{k_2}} & \cdots &H_{m,{k_K}} \end{array}\right],\qquad \qquad \textbf{z}_m \sim {\cal N}({\bf
	0},\bI).
\end{align*}
$\textbf{Q}$ is a real symmetric matrix that could be decomposed by EVD as $\bQ=\bU\boldsymbol{\Gamma}\bU^T$ and replaced into the first term of \eqref{UCGMBD2}, so we have
\begin{align}\label{UCGMBD21}
\xi_{m,n} & = \textbf{z}_m^T  \textbf{U}^T \boldsymbol{\Gamma}\underbrace{\textbf{U} \textbf{z}_m}_{\textbf{z}\sim{\cal N}({\bf
		0},\bI).} -  \mbox{Trace}(\boldsymbol{\Gamma}) = \sum_{i=1}^K\: \gamma_i z_i^2 - \sum_{i=1}^K\: \gamma_i
\end{align}
where $\gamma_i$ and $z_i$ are component of diagonal matrix $\boldsymbol{\Gamma}$ and vector $\bz$, respectively. Sum of independent chi-square random variables is well-known to be sub-exponential, and its moment generating function (MGF) can be bounded as
\begin{align}\label{UCGMBD22}
\textmd{E}\left[\exp\left(\lambda\xi_{m,n}\right)\right]\leq \prod_{i=1}^K \frac{e^{-\lambda\gamma_i}} {\sqrt{1 - 2\lambda \gamma_i}} \leq \prod_{i=1}^K e^{2\lambda^2 \gamma_i^2}
=e^\frac{4\lambda^2 \|\textbf{Q}\|_F^2}{2} \;; \quad |\lambda|< \frac{1}{4\|\textbf{Q}\|_2}
\end{align}
where the first and second inequalities were derived from \cite[Example 2.8]{wainwright_2019}. Now from \eqref{UCGMBD22} we compute $D_2$'s MGF as
\begin{align}\label{UCGMBD23}
\textmd{E}\left[e^{\lambda D_2}\right]&=\textmd{E}\left[e^{ \frac {\lambda} {M}\sum_{m=1}^M\xi_{m,n}}\right]=\prod_{m=1}^M\textmd{E}\left[e^{\frac {\lambda} {M}\xi_{m,n}}\right]\leq e^\frac{4\lambda^2 \|\textbf{Q}\|_F^2}{2M} \;; \quad |\lambda|< \frac{M}{4\|\textbf{Q}\|_2}
\end{align}
Note that given $\bc_k$'s and with $H_{m,n}$s as random variables $\xi_{m_1,n}$ and $\xi_{m_2,n}$ are independent for $m_1\neq m_2$ because channels across different antennas were assumed independent. Hence, we can take out the product from the expected value. Now, we have $\nu_2^2 = \frac{4\|\textbf{Q}\|_F^2}{M} $ and $b_2 = \frac{4\|\textbf{Q}\|_2}{M}$. For a more tractable appearance, norms of $\bQ$ should be upper bounded. We simplify them as follows
\begin{align*}\label{UCGMBD24}
\|\textbf{Q}\|_F^2 \leq 16 \big(\sum_{\substack{k\in \cS}} P_k\big)^2 \frac{\left(\log\log(L)^2\right)}{L^2},\qquad \|\textbf{Q}\|_2 \leq {\rm Trace}(\bQ)\leq 4 K P_{\max}\frac{\left(\log\log(L)\right)}{L},\qquad (a.s.)
\end{align*}
Concluding, we have
\begin{equation} \label{d_2}
\nu_2^2 = \frac{64K^2P_{\max}^2 \left(\log\log(L)\right)^2}{ML^2}, \qquad b_2 = \frac{16 K P_{\max}\log\log(L)}{ML}.
\end{equation}
Finally, we focus on $D_3$,
\begin{align}\label{UCGMBD3}
D_3 & = \frac{1}{M}\sum_{m=1}^M\: \Bigg[\left( \sum_{\substack{k\in \cS}} \sqrt{P_k} H_{m,k} (\bc_k^T\bc_n)\right)^2 +  \left(\bw_m^T c_n\right)^2 + 2\left(\bw_m^T c_n\right)\left(\sum_{\substack{k\in \cS}} \sqrt{P_k} H_{m,k} (\bc_k^T\bc_n)\right)\nonumber\\
&\quad-\left(\sum_{\substack{k\in \cS}} \sqrt{P_k} H_{m,k} (\bc_k^T\bc_n)\right)^2 -  \sigma_W^2\Bigg]= \frac{1}{M}\sum_{m=1}^M\: \underbrace{(f_m^2 + 2\alpha_{mn}f_m -\sigma_W^2) }_{g_{mn}}
\end{align}
where we have applied the change of variables $f_m=\bw_m^T\bc_n$ and $\sum_{\substack{k\in \cS}} \sqrt{P_k} H_{m,k} (\bc_k^T\bc_n) = \alpha_{mn}$. Note that with $\bC,\bH$ fixed and $\bw_m$ as random variables, $f_m$ is a linear combination of independent Gaussians and hence Gaussian itself. Indeed, $f_m\sim \cN(0,\sigma_W^2)$. Furthermore, $g_{m,n}=(f_m+\alpha_{m,n})^2-\sigma_W^2-\alpha_{m,n}^2$ is distributed as non-centralized chi-squared. Also note that $f_m$s are independent for different $m$ because noise is assumed independent across antennas. First, we derive the MGF for $g_{m,n}$ and then upper bound it using the result of\cite[Example 2.8]{wainwright_2019}
\begin{align}\label{UCGMBD32}
\textmd{E}\left[e^{\lambda g_{m,n}}|\textbf{C},\textbf{H}\right]&= \frac{e^{-\lambda\sigma_W^2-\lambda\alpha_{m,n}^2}} {\sqrt{1 - 2\lambda \sigma_W^2}}\times e^{\frac{\alpha_{m,n}^2\lambda}{1 - 2\lambda \sigma_W^2}} \leq e^{2\lambda^2 \sigma_W^4 + 4\alpha_{mn}^2\lambda^2\sigma_W^2}, \qquad|\lambda| <\frac{1}{4\sigma_W^2}
\end{align}
Next, MGF bound for $D_3$ is derived as
\begin{align}\label{UCGMBD33}
\textmd{E}\left[e^{\lambda D_3}\right]&=\textmd{E}\left[e^{ \frac {\lambda} {M}\sum_{m=1}^M g_{m,n}}\right]=\nonumber
\prod_{m=1}^M\textmd{E}\left[e^{\frac {\lambda} {M} g_{m,n}}\right]= e^{\frac{\lambda^2 2\sigma_W^4}{M} +\frac{4\lambda^2\sigma_W^2}{M^2}\sum_{m=1}^{M} \alpha_{m,n}^2},\qquad|\lambda| <\frac{M}{4\sigma_W^2}
\end{align}
where for large $M$ values $\frac{1}{M}\sum_{m=1}^M \alpha_{m,n}^2 \longrightarrow \textmd{E}(\alpha_{m,n}^2)$ almost surely by the strong law of large numbers. Note that expected value is taken over $\bH$ with $\bC$ fixed. We evaluate this expected value and bound the result with \eqref{random_walk}. After some simple algebra, our sub-exponential parameters are
\begin{align}
\nu_3^2 &= \frac{4\sigma_W^4 +\: 32\:KP_{\max}\sigma_W^2 \frac{\log\log(L)} {L}}{M},\qquad b_3 = \frac {4\sigma_W^2} {M}
\end{align}
For PoF analysis, we return to \eqref{prob_nocsi} and obtain \cite[Theorem 2.19]{wainwright_2019}:
\begin{align}\label{pof1OTD}
p(\theta_n - \textmd{E}(\theta_n) \geq t/2) = p\left(\sum_{i=1}^3 D_i \geq t/2\right) \leq
\left\{
\begin{array}{rl}
e^{-\frac{t^2}{4\nu_{\ast}^2}} & \text{if }\; 0\leq t/2 \leq \frac{\nu_{\ast}^2}{b_{\ast}} ,\\
e^{-\frac{t}{2 b_{\ast}}} & \text{if }\;  t/2 > \frac{\nu_{\ast}^2}{b_{\ast}}.
\end{array} \right.
\end{align}
where $\ast=nb$ and $\nu_{nb}^2 = \nu_1^2 + \nu_2^2 + \nu_3^2$, $b_{nb} =\max(b_1 , b_2 , b_3)$ and $t =P_{\min}(1-\frac{1}{L}) $. Note that $nb$ stands for `N'o CSI, and `B'ad statistics.

\subsection{Active users without CSI}
Firstly, we have
\begin{equation} \label{prob_nocsig}
p\left(\theta_n \leq \rho_1\right)=p\left(\theta_n-{\rm E}[\theta_n]\leq\rho_1-{\rm E}[\theta_n]\right)\leq p\left(\theta_n-{\rm E}[\theta_n]\leq \frac{t}{2}\right)
\end{equation}
For active users, we separate $\{H_{m,n}\}_{m=1}^M$ from $\bH$, therefore, the Martingale difference sequence (MDS) has four terms.
\begin{eqnarray} \label{mds}
\theta_n-{\rm E}[\theta_n]&=&\left(\theta_n-{\rm E}[\theta_n|\bC,\bH]\right)+\left({\rm E}[\theta_n|\bC,\bH]-{\rm E}[\theta_n|\bC,\bH-\{H_{m,n}\}_{m=1}^{M}]\right)\nonumber\\ &&+\left({\rm E}[\theta_n|\bC,\bH-\{H_{m,n}\}_{m=1}^{M}]-{\rm E}[\theta_n|\bC]\right)+\left({\rm E}[\theta_n|\bC]-{\rm E}[\theta_n]\right)\nonumber \\ &:=&D_4+D_3+D_2+D_1
\end{eqnarray}
The MGF bounds for $D_1,D_2$ are almost similar to case of inactive users without CSI and are omitted. The outcomes are the following equations
\begin{eqnarray} \label{active}
\nu_1^2=\frac{4(K-1)P_{\max}^2}{L^2},&\qquad & b_1=\frac{8(K-1)P_{\max}\log\log(L)}{L}\\ \nu_2^2=\frac{64(K-1)^2P_{\max}^2(\log\log(L))^2}{ML^2}, &\qquad & b_2=\frac{16(K-1)P_{\max}\log\log(L)}{ML} 
\end{eqnarray}
The term $D_3$ did not exist in the inactive users case, and is a new term. Hence, we derive its bound in detail. First, note that by the change of variables $\sum_{\substack{k\in \cS\\k\neq n}} \sqrt{P_k} H_{m,k} (\bc_k^T\bc_n) = \beta_{m,n}$, we rewrite $D_3$ as
\begin{align}\label{UCGMGD31}
D_3 & = \frac{1}{M}\sum_{m=1}^M\: \underbrace{(\sqrt{P_n} H_{m,n} + \beta_{m,n}) ^2}_{g_{m,n}}- \underbrace{(\beta_{m,n}^2 + P_n)}_{\textmd{E}\left[g_{m,n}\right]}
\end{align}
Note that we have a different definition for $g_{m,n}$ compared to the inactive users' case which is a simple abuse of notation to avoid notation explosion. All the terms we have bounded till now, used two-sided MGF bounds. We note that for \eqref{prob_nocsig}, one sided bounds from below are enough. This significantly simplifies our derivations. Using the one-sided Bernstein-type bound in \cite[Proposition 2.14]{wainwright_2019}, we have
\begin{align}
\textmd{E}\left[e^{\lambda\left(- g_{m,n} +   \textmd{E}\left[g_{m,n}\right]\right)}\Big|\textbf{C},\bH-\{H_{m,n}\}_{m=1}^{M}\right]&\leq e^{\frac{\frac{\lambda^2}{2}\textmd{E}\left[g_{m,n}^2\big|\textbf{C},\bH-\{H_{m,n}\}_{m=1}^{M}\right]}{1 -  \frac{b \lambda}{3}}} \: ;\lambda \in [0,\frac{3}{b})
\end{align}
where $\lambda \geq 0$ , $(-g_{m,n} \leq b) $ and $b = 0$. The expectation term in the exponential argument can be evaluated as $\textmd{E}\left[g_{m,n}^2\big|\textbf{C},\bH-\{H_{m,n}\}_{m=1}^{M}\right] = 3 \: P_n^2 + \beta_{m,n}^4 + 6 \: P_n \: \beta_{m,n}^2
$. Plugging $D_3$ from \eqref{UCGMGD31}, we can bound its MGF as follows
\begin{align}\label{UCGMGD33}
\textmd{E}\left[e^{\lambda D_3}\big|\textbf{C},\bH-\{H_{m,n}\}_{m=1}^{M}\right]
&\leq e^{\frac{\frac{\lambda^2}{2} \left(\frac{1}{M}\sum_{m=1}^M\: \:\textmd{E}\left(g_{m,n}^2\big|\textbf{C},\bH-\{H_{m,n}\}_{m=1}^{M}\right)\right)}{M}}  ;\lambda \in (- \infty , 0]
\end{align}
Note that while $P_n$ is deterministic, $\beta_{m,n}$ is a random variable defined as a function of $H_{m,k},k\neq n,\bC$. Due to channel independence, $\beta_{m,n}$ are independent across $m$. Thus, we invoke strong law of large numbers to deduce that $\frac{1}{M}\sum_{m=1}^M\: \:\textmd{E}\left(g_{m,n}^2\big|\textbf{C},\bH-\{H_{m,n}\}_{m=1}^{M}\right)$ converges to $\textmd{E}\left(g_{m,n}^2\big|\textbf{C}\right) $ almost surely.
Assuming the almost sure bound of Radmacher codes in \eqref{random_walk} we have our sub-Gaussian parameters as
\begin{align}
\nu_3^2 &= \frac{1}{M}\Big(3 P_{\max}^2 + 48 (K-1)^2P_{\max}^2 \frac{\left(\log\log(L)\right)^2}{L^2} + 24 (K-1)P_{\max}^2 \frac{\left(\log\log(L)\right)}{L}\Big),\; b_3 = 0
\end{align}
Finally, the evaluation of $D_4$ is almost the same as $D_3$ for the inactive users case and yields the following sub-exponential parameters
\begin{align}
\nu_4^2 &= \frac{4\sigma_W^4 + 8 \:\sigma_W^2 P_{\max}  + \: 32\:\sigma_W^2 (K-1)P_{\max} \frac{\log\log(L)}{L}}{M}, \qquad b_4 = \frac {4\sigma_W^2} {M}
\end{align}
PoF bound for $p(\theta_n - \textmd{E}(\theta_n) \leq t/2)$ is derived as in \eqref{pof1OTD} with $\ast=ng$ where $\nu_{ng}^2 = \nu_1^2 + \nu_2^2 + \nu_3^2 + \nu_4^2$, $b_{ng} = \max(b_1 , b_2 , b_3 , b_4 )$. Note that $ng$ stands for `N'o CSI, and `G'ood statistics.

\subsection{Inactive users with CSI}
The corresponding probability can be simplified as in \eqref{prob_nocsi}, while the corresponding MDS is written similar to \eqref{mds}.
Utilizing the same techniques that we applied to inactive users with no CSI $D_1,D_2$ MGFs can be bounded as sub-exponential random variables with the same parameters as \eqref{varD1BUCGM}, \eqref{bD1BUCGM}, and \eqref{d_2}. Let us focus on $D_3$ next,
\begin{align}\label{NCGMBD3}
D_3 & = \frac{1}{M}\sum_{m=1}^M\: (H_{m,n}^2-1)\left[\left(\sum_{\substack{{k}\in \cS}} \sqrt{P_{k}} H_{m,{k}}(\bc_{k}^T\bc_n)\right)^2 + \sigma_W^2\right]:=\frac{1}{M}\sum_{m=1}^M\: \left[ \underbrace{\beta_{m,n} H_{m,n}^2}_{g_{m,n}}- \underbrace{\beta_{m,n}}_{\textmd{E}\left[g_{m,n}\right]}\right]
\end{align}
where $\beta_{m,n}$ is defined accordingly. Again, by exploiting MGF bound for a centralized chi-squared variable \cite[Example 2.8]{wainwright_2019}, we have
\begin{align}\label{NCGMBD32}
\textmd{E}\left[e^{\lambda\left(g_{m,n} -   \textmd{E}\left[g_{m,n}\right]\right)}\Big|\textbf{C},\bH-\{H_{m,n}\}_{m=1}^M\right]&= e^{2\lambda^2\beta_{m,n}^2},\qquad \lambda \leq \frac{1}{4\beta_{m,n}}
\end{align}
Utilizing independence of $\beta_{m,n}$ across antennas, we arrive at
\begin{align}\label{NCGMBD33}
\textmd{E}\left[e^{\lambda D_3}\big|\textbf{C},\bH-\{H_{m,n}\}_{m=1}^M\right]
&\leq e^{\frac{2\lambda^2 \left(\frac{1}{M}\sum_{m=1}^M\:\beta_{m,n}^2\right)}{M}}
&\leq  e^{\frac{2\lambda^2  \textmd{E}\left(\beta_{m,n}^2\right)}{M}},\qquad \lambda \leq \frac{1}{\frac{4 \max_{1 \leq m \leq M}(\beta_{m,n})}{M}}
\end{align}
Invoking strong law of large numbers (SLLN), we have $1/M\sum_{m=1}^M \beta_{m,n}^2\longrightarrow E[\beta_{m,n}^2]$ almost surely. Furthermore, we can write
\[
p\left(\bigcup_{m=1}^{M} \left\{\beta_{m,n}>\tau_0\log M\right\}\right) \leq \sum_{m=1}^{M}p(\beta_{m,n}>\tau_0\log M) \leq \sum_{m=1}^{M}e^{-\frac{\tau_0\log M-{\rm E}[\beta_{m,n}]}{2b}} \leq e^{\frac{{\rm E}[\beta_{m,n}]}{2b}} <\infty
\]
We have used union bound in the first inequality, sub-exponential property of chi-squared in the second inequality, and the assumption $\tau_0=2b$ for the third inequality. Next, we apply Borel-Cantelli Lemma \cite{papoulis02} and deduce that since the right hand side is finite, then almost surely a finite number of $\beta_{m,n}>\tau_0\log M$ happen. Let us call the maximum of these finite violations by $\tau_{\max}$ which is finite. Then,
we have
\begin{equation} \label{B-C}
4\frac{\underset{1\leq m\leq M}{\max}~\beta_{m,n}}{M} \leq \frac{4\max\{\tau_{\max},\tau_0\log M\}}{M}~\underset{M\rightarrow\infty}{\longrightarrow} ~ 0 \qquad (a.s.)
\end{equation}
Thus, using \eqref{NCGMBD33} and \eqref{B-C}, MGF for $D_3$ is bounded with sub-exponential parameters
\begin{align}
\nu_3^2 &= \frac{4\sigma_W^4 + 192 K^2P_{\max}^2 \frac{\left(\log\log(L)\right)^2}{L^2} + 32 \: \sigma_W^2 KP_{\max} \frac{\left(\log\log(L)\right)}{L}}{M}, \qquad b_3 = 0
\end{align}
Finally, we evaluate $D_4$
\begin{align}\label{NCGMBD4}
D_4 =\frac{1}{M}\sum_{m=1}^M\: \underbrace{H_{m,n}^2(f_m^2 + 2\alpha_{m,n}f_m -\sigma_W^2) }_{g_{mn}}
\end{align}
where we have defined $f_m:=w_m^T\bc_n$ and $\alpha_{m,n}:=\sum_{\substack{k\in \cS}} \sqrt(P_k) H_{m,k} (\bc_k^T\bc_n)$,
Utilizing the MGF bound for non-centralized chi-squared, we have
\begin{align}\label{NCGMBD42}
\textmd{E}\left[e^{g_{m,n}}|\textbf{C},\textbf{H}\right]
&\leq e^{\frac{\lambda^2 H_{m,n}^4(4\sigma_W^4 + 8\alpha_{mn}^2\sigma_W^2)}{2}}  , \qquad|\lambda| <\frac{1}{4\sigma_W^2 H_{m,n}^2}
\end{align}
Similarly, for $D_4$, we have
\begin{align}\label{NCGMBD43}
\textmd{E}\left[e^{\lambda D_4}\right]&=\textmd{E}\left[e^{ \frac {\lambda} {M}\sum_{m=1}^M g_{m,n}}\right]
\leq e^{\frac{\lambda^2 \left[4\sigma_W^4 \left (\frac{1}{M}\sum_{m=1}^M H_{m,n}^4\right) +\: 8\:\sigma_W^2\left(\frac{1}{M}\sum_{m=1}^M H_{m,n}^4\alpha_{m,n}^2\right)\right]}{2M}}  ;|\lambda| < \frac{1}{\frac{4\sigma_W^2 \max_{1 \leq m \leq M}(H_{m,n}^2)}{M}}
\end{align}
Invoking independence over $m$ and SLLN, we almost surely have $\frac{1}{M}\sum_{m=1}^M H_{m,n}^4 \longrightarrow \textmd{E}(H_{m,n}^4 )$, and  $\frac{1}{M}\sum_{m=1}^M H_{m,n}^4\alpha_{m,n}^2 \longrightarrow \textmd{E}(\alpha_{m,n}^2)\textmd{E}(H_{m,n}^4)$. Furthermore, using the same process as in \eqref{B-C} and applying Borel-Cantelli Lemma, we can bound the $\max$ in \eqref{NCGMBD43}. Hence, sub-Gaussian parameters are given by
\begin{align}
\nu_4^2 &= \frac{4\sigma_W^4 + \: 96\:\sigma_W^2 \frac{\log\log(L)KP_{\max}}{L}}{M}, \qquad
b_4 = 0
\end{align}
Finally, we have $\ast=fb$ in \eqref{pof1OTD} where $\nu_{fb}^2 = \nu_1^2 + \nu_2^2 + \nu_3^2 + \nu_4^2$, $b_{fb} = \max(b_1 , b_2 , b_3 , b_4) $. Note that $fb$ stands for `F'ull CSI, and `B'ad statistics.

\subsection{Active users with CSI}
Again, we simplify the probability as in \eqref{prob_nocsig} but with $t$ replaced by $t_f:=(3-1/L)P_{\min}$. Note that while we could have replaced $t_f$ with $t$ and still obtained a valid result, $t_f$ yields a smaller error and thus a tighter bound. Then, we write the corresponding MDS as in \eqref{mds}. Using the same techniques, we deduce that $D_1,D_2$ are sub-exponential with parameters given by \eqref{active}.
For $D_3$, we apply the one-sided tail bound that we exploited in active users without CSI and obtain
\begin{align}
\nu_3^2 &= \frac{\left[105 \: P_{\max}^2 + 144 (K-1)^2P_{\max}^2 \frac{\left(\log\log(L)\right)^2}{L^2} + \Big(360 \: P_{\max} + 24 \: \sigma_W^2\Big)(K-1)P_{\max} \frac{\left(\log\log(L)\right)}{L}\right]}{M}\nonumber \\&\qquad + \frac{3\sigma_W^4 + 30 \: P_{\max} \sigma_W^2}{M},\qquad b_3 = 0
\end{align}
Then, we apply the same technique as in inactive users with CSI to $D_4$ to obtain
\begin{align}
\nu_4^2 &= \frac{12\sigma_W^4 + \: 8\:\sigma_W^2 \left(15 P_{\max} + 12 (K-1)P_{\max} \frac{\log\log(L)}{L}\right)}{M},\qquad
b_4 = 0
\end{align}
Finally, we get $p(\theta_n - \textmd{E}(\theta_n) \leq t_f/2)$ bounded as in \eqref{pof1OTD} with $\ast=fg$ where $\nu_{fg}^2=\nu_1^2+\nu_2^2+\nu_3^2+\nu_4^2$ and $b_{fg}=\max(b_1,b_2,b_3,b_4)$ and $t$ replaced by $t_f$. Note that $fg$ stands for `F'ull CSI, and `G'ood statistics.

\subsection{Combining the results}
Plugging from \eqref{pof1OTD} for the four different cases into \eqref{PoF}, we obtain
\begin{align}
\textmd{PoF} \leq \label{OTD_POFME}
(N -|\Lambda|-K_2) \left\{
\begin{array}{rl}
e^{-\frac{t^2}{4\nu_{nb}^2}} & \text{if }\; 0\leq t/2 \leq \frac{\nu_{nb}^2}{b_{nb}}  ,\\
e^{-\frac{t}{2 b_{nb}}} & \text{if }\;  t/2 > \frac{\nu_{nb}^2}{b_{nb}}\;.
\end{array} \right. + (K_2) \left\{
\begin{array}{rl}
e^{-\frac{t^2}{4\nu_{ng}^2}} & \text{if }\; 0\leq t/2 \leq \frac{\nu_{ng}^2}{b_{ng}},\\
e^{-\frac{t}{2 b_{ng}}} & \text{if }\;  t/2 > \frac{\nu_{ng}^2}{b_{ng}}\;.
\end{array} \right.\nonumber\\
+ (|\Lambda|-K_1) \left\{
\begin{array}{rl}
e^{-\frac{t^2}{4\nu_{fb}^2}} & \text{if }\; 0\leq t/2 \leq \frac{\nu_{fb}^2}{b_{fb}},\\
e^{-\frac{t}{2 b_{fb}}} & \text{if }\;  t/2 > \frac{\nu_{fb}^2}{b_{fb}}\;.
\end{array} \right. + (K_1) \left\{
\begin{array}{rl}
e^{-\frac{t_f^2}{4\nu_{fg}^2}} & \text{if }\; 0\leq t_f/2 \leq \frac{\nu_{fg}^2}{b_{fg}},\\
e^{-\frac{t_f}{2 b_{fg}}} & \text{if }\;  t_f/2 > \frac{\nu_{fg}^2}{b_{fg}}.
\end{array} \right.
\end{align}
Here, $K_1$ is the number of active users with CSI and $K_2$ is the number of active users without CSI. Finally, we use a tractable tight upper bound of \eqref{OTD_POFME} by employing the well-known log-sum-exp inequality of the following Lemma.\\
{\bf Lemma 2.} For any real numbers $x_1,\dots,x_n$, we have
\[
\exp(\max\{x_1,\dots,x_n\})\leq \sum_{i=1}^ne^{x_i}\leq n\exp(\max\{x_1,\dots,x_n\}).
\]
By some simple manipulations of \eqref{OTD_POFME} according to Lemma 2, we obtain the final result. After further simplifications, that are carried out in the performance analysis section, we realized that in most practical scenarios the lower branch conditions in \eqref{OTD_POFME} are satisfied. Therefore, we drop the upper branch and this concludes our derivation of PoF for OMC.

\section*{Appendix B. Derivation of PoF for OTD}
Applying union bound, PoF is bounded as
\begin{eqnarray*}
\textrm{PoF} & \leq & p\left(\min_{n\in \mathcal{S}\cap \bar{\Lambda}}\:\theta_n \leq T_{\bar{\Lambda}}\right)+ p\left( \max_{n\in  \bar{\mathcal{S}}\cap\bar{\Lambda} } \:\theta_n \geq T_{\bar{\Lambda}}\right) +p\left(\min_{n\in \mathcal{S}\cap \Lambda}\:\theta_n \leq T_{\Lambda}\right)+ p\left( \max_{n\in  \bar{\mathcal{S}}\cap\Lambda } \:\theta_n \geq T_{\Lambda}\right) 
\end{eqnarray*}
We can use the bounds for OMC to bound the PoF for OTD also. Note that the terms on the right hand side in the four probabilities above are different from OMC as two thresholds are utilized in OTD compared to picking the $K$ maximum values in OMC. Dropping the first branches in \eqref{OTD_POFME} because their conditions are not valid most of the time, we arrive at PoF for OTD.

% or


%\ifCLASSOPTIONcaptionsoff
%  \newpage
%\fi
%\newpage




\bibliographystyle{IEEEtran}
\bibliography{otd2_ref}

\end{document}